%% file: main.tex
\documentclass[10pt,letterpaper,onecolumn,draftclsnofoot]{IEEEtran}

\usepackage{tikz}
\tikzstyle{arrow} = [thick,->,>=stealth]
\usetikzlibrary{decorations.pathreplacing}
\usepackage[cmex10]{amsmath}
\usepackage{amsthm}
\usepackage{amssymb,amsfonts,dsfont,bm,mathrsfs}
\usepackage{siunitx}
\usepackage[bookmarksopen=true,pagebackref=true]{hyperref}
\usepackage{cleveref}
\usepackage[noadjust]{cite}
\usepackage{graphicx}
\allowdisplaybreaks[2]
\usepackage{stfloats}
\usepackage{enumerate}
\usepackage{makecell}
\usepackage{multirow}
\usepackage{algorithmic}
\usepackage{algorithm}

{
\theoremstyle{plain}
\newtheorem{theorem}{Theorem}

\newtheorem{lemma}{Lemma}
\newtheorem{corollary}{Corollary}

\theoremstyle{definition}
\newtheorem{definition}{Definition}
\newtheorem{example}{Example}

\theoremstyle{remark}
\newtheorem{remark}{Remark}
}

\begin{document}

\title{Distributed Approximate Computing with Constant Locality}

%

\author{Deheng Yuan, Tao Guo, Zhongyi Huang and Shi Jin
\thanks{Deheng Yuan and Zhongyi Huang are with the Department of Mathematical Sciences, Tsinghua University, Beijing 100084, China (emails: ydh22@mails.tsinghua.edu.cn, zhongyih@tsinghua.edu.cn).}
\thanks{Tao Guo is with the School of Cyber Science and Engineering, Southeast University, Nanjing 210096, China (email: taoguo@seu.edu.cn).}
\thanks{Shi Jin is with the National Mobile Communications Research Laboratory, Southeast University, Nanjing, 210096, China (e-mail: jinshi@seu.edu.cn).}
}

\maketitle



\ifCLASSOPTIONpeerreview
\begin{center} \bfseries EDICS Category: 3-BBND \end{center}
\fi
%

\begin{abstract}
Consider a distributed coding for computing problem with constant decoding locality, i.e., with a vanishing error probability, any single sample of the function can be approximately recovered by probing only constant number of compressed bits.
We establish an achievable rate region by designing an efficient layered coding scheme, where the coding rate is reduced by introducing auxiliary random variables and local decoding is achieved by exploiting the expander graph code.  
Then we show the rate region is optimal under mild regularity conditions on source distributions. 
The proof relies on the reverse hypercontractivity and a rounding technique to construct auxiliary random variables. 
The rate region is strictly smaller than that for the classical problem without the constant locality constraint in most cases, which indicates that more rate is required in order to achieve lower coding complexity.
Moreover, a coding for computing problem with side information is analogously studied.
We also develop graph characterizations, which simplifies the computation of the achievable rate region.
\end{abstract}


\IEEEpeerreviewmaketitle

\input{1}

\input{2}
\input{3}

\input{4}

\input{5}
\input{6}

\input{7}
\input{8}
\input{A1}

\input{A2}
\input{A3}
\input{A4}

\input{A5}
\input{A6}

\bibliographystyle{bibliography/IEEEtran}
\bibliography{bibliography/computing}






\end{document}

%% file: 1.tex
\section{Introduction}
Nowadays, many applications depend on the efficiency of distributed computing by cooperative agents, such as in scenarios of cloud computing~\cite{Soyata2012}, distributed optimization~\cite{Nedic2009} and machine learning~\cite{Lee2018}.
To save resources, massive data has to be compressed such that computing tasks can be handled at a fidelity level with a rate as low as possible.
Moreover, fast decoding and reconstruction are also of great importance, especially for latency sensitive applications.

From an information-theoretic point of view, the compression limits for computing a function were studied as the coding for computing problems in~\cite{Yamamoto1982,Orlitsky2001,Han1987}. 
They extended earlier works~\cite{SlepianWolf1973,WynerZiv1976,Berger1978,Tung1978,Wagner20081} on distributed compression problems, where the identity function is computed. 
The optimal rate for the problem where decoder side information is available were fully solved in~\cite{WynerZiv1976,Yamamoto1982,Orlitsky2001}.
While for the distributed computing problem, the optimal rate region remains unclear, and special cases were fully solved in~\cite{Han1987,KornerMarton1979,Wagner20082}.

For many applications,  lossless computing incurs a high cost. And an alternative is to impose a relaxed approximate computing constraint~\cite{Xu2016}. 
The coding for approximate computing problem was studied, first for single source~\cite{Posner1971} and then for distributed sources~\cite{Basu2020,Basu2022}.
The constraint was stronger than the average distortion constraint in usual rate-distortion problems. 
The resulting problem is often easier to handle compared to problems that require the strong converse~\cite{Marton1974}.

The rate regions for many problems were characterized in terms of auxiliary random variables and implicit reconstruction functions, which makes the computation difficult. Graph-based characterizations were developed for various problems~\cite{Orlitsky2001,Doshi2010,Basu2022,Yuan2022,Yuan2023} to handle the computation.
The meaning of the random variables was explicitly described by graphs and the reconstruction function is implicit in the construction of edges. This made the computation of the rate regions much easier both analytically and numerically.

The tradeoff between the rate and the reconstruction quality was the main concern of these theoretic frameworks, while in contrast, reconstruction complexity was not taken into consideration. 
In~\cite{Makhdoumi2015,Mazumdar2014}, the number of accessed bits for decoding each source symbol (defined as the decoding locality) was used  as a measure of such complexity.
Note that the decoding locality is different from related notions of updating locality (the number of compressed bits influenced by some source symbol) and encoding locality (the number of source symbols 
influencing some compressed bit) studied by many works~\cite{Montanari2008,Mazumdar2014,Vatedka2020,Mazumdar2021}.
It is shown in~\cite{Makhdoumi2015} that for the compression of one source, a decoding locality of $O(\log n)$ can be achieved without loss of rate by the simple concatenated coding, where $n$ is the source length.
It was further shown in~\cite{Mazumdar2015,Pananjady2018,Tatwawadi2018}, by exploiting the expander graph code for storing sparse bit strings~\cite{Buhrman2002}, that constant decoding locality can be achieved without incurring an excess rate.

However, for the distributed lossless compression problem with constant decoding locality in~\cite{Vatedka2022}, the classical Slepian-Wolf region is not always achievable.
In particular, for confusable sources defined therein, the compressed rate for each source has to be larger than the source entropy. 
For non-confusable sources, it was proved that  at least one of the optimal rates must be strictly smaller than the entropy. 
However, a coding scheme achieving the rates was not given for non-confusable sources. 

In the current work, we consider the coding for distributed approximate computing problem with constant decoding locality\footnote{In this work, we restrict our scope to the decoding locality, and without ambiguity we refer to it as locality.}. 
First we design a layered coding scheme, which induces an achievable rate region for general sources.
The scheme takes advantages of both joint typicality coding techniques and the expander graph code.
Each source sequence is first divided into blocks, and each block is encoded by typicality coding. Hence auxiliary random variables can be introduced in the coding of each block to eliminate the redundancy for computing the function.
Then the blocks where the typicality coding fails are encoded by the expander graph code to ensure constant decoding locality.
The induced rate region can be specialized to establish an achievable region for the distributed lossless compression problem in~\cite{Vatedka2022}.

Then we show the layered coding scheme is optimal for sources with a full support, which implies that the optimal rate region for our problem can be strictly smaller than that for the problem without constant locality. 
The proof mainly relies on the reverse hypercontractivity property and a rounding technique to construct auxiliary random variables.
By utilizing the reverse hypercontractivity property, constraints on the joint distribution are transformed to that on marginal distributions of different encoder sides. 
Then we apply rounding techniques to the marginal distributions to construct auxiliary random variables. 
This induces an outer bound that coincides with the achievable region, and thus proves the optimality.
Moreover, we show that the optimality remains true if the approximate computing constraint is replaced by a vanishing average distortion constraint, or the constant locality constraint is weakened to a bounded average locality constraint. 
This implies that the crucial reason for the rate region to get smaller is the intrinsic low complexity of the decoding function. That is,  more rate is necessary in order to achieve lower coding complexity. 
We further generalize both the achievable and converse parts to the distributed computing problem with more than two encoders.

Next, the analogous coding for computing problem with side information is considered. On one hand, a similar layered coding scheme is easily obtained from that for the distributed problem. On the other hand, the converse is proved by a coupling technique and its induced virtual distributed coding problem. 

Finally, we develop graph characterizations for the above rate regions. For the distributed computing problem, the region is shown to be an infinite polygon determined by finite number of rate pairs, each of which corresponds to a distributed characteristic bipartite graph. Then the rate for each side can be independently optimized. The resulting optimization problem is a special case of the source-channel optimization problem in~\cite{Yuan2023}. 
Hence the computation of the rate region becomes easier both analytically and numerically. 

The rest of the paper is organized as follows. The problem formulations and preliminaries are presented in
\Cref{sec:PS}. Then we establish the achievable rate region in~\Cref{sec:achievable} and prove its converse in~\Cref{sec:converse}. In~\Cref{sec:more}, results are generalized to the case with more than two encoders. The approximate computing problem with side information is solved in~\Cref{sec:sideinformation}.
Then we develop graph characterizations and present several examples in \Cref{sec:graph}. Finally, we conclude our work in \Cref{sec:conclusion}.

%% file: 2.tex
\section{Problem Formulation and Preliminaries}
\label{sec:PS}
\subsection{Local Decoding in Distributed Approximate Computing}
\label{subsec:distributed}
Denote a discrete random variable by a capital letter and its finite alphabet by the corresponding calligraphic letter, e.g., $X_1\in\mathcal{X}_1$ and $\hat{Z}\in\hat{\mathcal{Z}}$.
We use the superscript $n$ to denote an $n$-sequence, e.g., $X^n=(X_{i})_{i = 1}^n$.

Let $(X_{1i},X_{2i}) \sim p(x_1,x_2),i\in\{1,2,\cdots,n\}$ be i.i.d. random variables distributed over $\mathcal{X}_1\times\mathcal{X}_2$. Without loss of generality, assume $p(x_1) > 0$ and $p(x_2)>0$, $\forall x_1 \in \mathcal{X}_1$, $x_2 \in \mathcal{X}_2$.

\begin{figure}[!t]
\centering
\begin{tikzpicture}
	\node at (0.5,0.8) {$X_1^n$};
	\draw[->,>=stealth] (0.8,0.8)--(1.5,0.8);
	
	\node at (2.3,0.8) {Encoder $1$};
	\draw (1.5,0.4) rectangle (3.1,1.2);
	
	\draw[->,>=stealth] (3.1,0.8)--(4.7,0.8)--(4.7,0.4);
	\node at (3.9,1.0) {$M_1^{nR_1}$};

        \node at (0.5,-0.9) {$X_2^n$};
	\draw[->,>=stealth] (0.8,-0.9)--(1.5,-0.9);
	
	\node at (2.3,-0.9) {Encoder $2$};
	\draw (1.5,-1.3) rectangle (3.1,-0.5);
	
	\draw[->,>=stealth] (3.1,-0.9)--(4.7,-0.9)--(4.7,-0.4);
	\node at (3.9,-0.65) {$M_2^{nR_2}$};
	
	\node at (4.75,0) {Decoder};
	\draw (4.0,-0.4) rectangle (5.5,0.4);
	
	\draw[->,>=stealth] (5.5,0)--(6.0,0);
	\node [right] at (6.0,0) {$\hat{Z}^n (\epsilon)$};
\end{tikzpicture}
\caption{Distributed approximate computing, where $\hat{Z}_i = \mathrm{Dec}_i(M_{1,I_{1,i}},M_{2,I_{2,i}})$.}
\label{fig:system_model2}
\vspace{-0.2cm}
\end{figure}
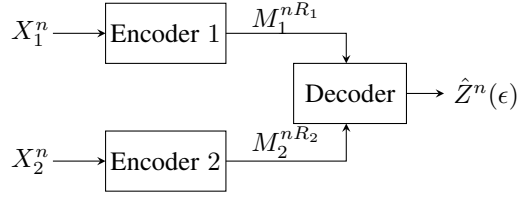

Consider the distributed approximate computing problem depicted in Fig.~\ref{fig:system_model2}.
The source messages $X_1^n$  and $X_2^n$ are observed by encoders 1 and 2, respectively.
The decoder needs to approximately compute a function $f : \mathcal{X}_1 \times \mathcal{X}_2 \to \mathcal{Z}$ within a certain distortion tolerance. Denote $f(X_{1i},X_{2i})$ by $Z_i$ for $1\leq i\leq n$. Let $d: \mathcal{Z} \times \hat{\mathcal{Z}} \to [0,\infty)$ be a distortion measure. Assume $\mathcal{Z} \subseteq \hat{\mathcal{Z}}$ and $d(z,z) = 0, \forall z \in \mathcal{Z}$.

An $(n,2^{nR_1},2^{nR_2},t)$ distributed code is defined by encoding functions 
\begin{align*}
\mathrm{Enc}_1: \mathcal{X}_1^n \to \{0,1\}^{nR_1},
\\
\mathrm{Enc}_2: \mathcal{X}_2^n \to \{0,1\}^{nR_2},
\end{align*}
a series of subsets $(I_{1,i})_{i =1}^{n}$ and $(I_{2,i})_{i =1}^{n}$ such that
\begin{equation}
\label{eq:worstlocality}
\begin{aligned}
    |I_{1,i}| \leq t, I_{1,i} \subseteq \{1,2,...,nR_1\}, \forall i \in\{1,...,n\},
    \\
    |I_{2,i}| \leq t, I_{2,i} \subseteq \{1,2,...,nR_2\}, \forall i \in\{1,...,n\},
\end{aligned}
\end{equation}
and a series of local decoding functions $(\mathrm{Dec}_i)_{i =1}^n$
\[
\mathrm{Dec}_i:\{0,1\}^{|I_{1,i}|} \times \{0,1\}^{|I_{2,i}|} \to \hat{\mathcal{Z}}, \forall i \in\{1,...,n\}.
\]
Then the encoded messages are binary sequences $M_1^{nR_1} = \mathrm{Enc}_1(X_1^n)$ and $M_2^{nR_2} = \mathrm{Enc}_2(X_2^n)$, and their components with indices in $I_{1,i}$ and $I_{2,i}$ are denoted by $M_{1,I_{1,i}}$ and $M_{2,I_{2,i}}$, respectively. The reconstruction for each $Z_i$ is $\hat{Z}_i = \mathrm{Dec}_i(M_{1,I_{1,i}},M_{2,I_{1,i}})$. Note that the decoding functions can be different for distinct symbols $Z_i$.  Moreover, the numbers of probed bits for all decoding functions are bounded by a constant $t$, which is called the locality.

A distributed rate-tolerance triple $(R_1,R_2,\epsilon)$ is called {\it  achievable} if there exists an $(n,2^{nR_1},2^{nR_2},t)$ distributed code such that 
\begin{equation}
\label{eq:distrachivable}
    \lim_{n \to \infty} \max_{1 \leq i \leq n} \mathbb{P}[d({Z}_i,\hat{Z}_i)> \epsilon] = 0.
\end{equation}
The rate region $\mathscr{R}_{loc}(\epsilon)$ for distributed approximate computing with local decoding constraints is defined as the closure  of the set of all the rate pairs $(R_1,R_2)$ such that $(R_1,R_2,\epsilon)$ is achievable.

\subsection{Local Decoding in Approximate Computing with Side Information}
\label{subsec:sideinformation}

Let $(S_{1i},S_{2i}) \sim p(s_1,s_2),i\in\{1,2,\cdots,n\}$ be i.i.d. random variables distributed over $\mathcal{S}_1\times\mathcal{S}_2$. 
Without loss of generality, assume $p(s_1) > 0$ and $p(s_2)>0$, $\forall s_1 \in \mathcal{S}_1$, $s_2 \in \mathcal{S}_2$.

\begin{figure}[!t]
\centering
\begin{tikzpicture}
	\node at (0.5,0) {$S^n$};
	\draw[->,>=stealth] (0.8,0)--(1.5,0);

    \node at (2.25,-1.25) {$\tilde{S}_1^n$};
	\draw[->,>=stealth] (2.25,-1.0)--(2.25,-0.4);
 
    \node at (4.75,-1.25) {$S_2^n$};
	\draw[->,>=stealth] (4.75,-1.0)--(4.75,-0.4);
	
	\node at (2.25,0) {Encoder};
	\draw (1.5,-0.4) rectangle (3.0,0.4);
	
	\draw[->,>=stealth] (3.0,0)--(4,0);
	\node at (3.5,0.2) {$M^{nR}$};
	
	\node at (4.75,0) {Decoder};
	\draw (4.0,-0.4) rectangle (5.5,0.4);
	
	\draw[->,>=stealth] (5.5,0)--(6.0,0);
	\node [right] at (6.0,0) {$\hat{Z}^n (\epsilon)$};
\end{tikzpicture}
\caption{Approximate computing with side information, where $S_1^n = (S^n,\tilde{S}_{1}^n)$ and $\hat{Z}_i = \mathrm{Dec}_i(M_{I_i},S_2^n)$.}
\label{fig:system_model1}
\vspace{-0.2cm}
\end{figure}
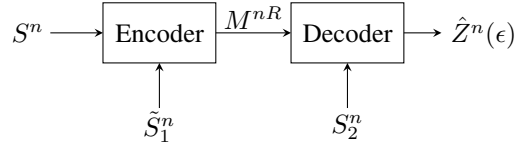

Consider the approximate computing problem with two-sided information depicted in Fig.~\ref{fig:system_model1}. The source message $S_1^n$ observed by the encoder has two parts $(S^n,\tilde{S}_{1}^n)$, with $S^n$ being the original source and $\tilde{S}_1^n$ being the encoder side information. 
The other part $S_2^n$ of the side information is observed by the decoder.
The decoder needs to approximately compute a function $f : \mathcal{S}_1 \times \mathcal{S}_2 \to \mathcal{Z}$ within a certain distortion tolerance. Denote $f(S_{1i},S_{2i})$ by $Z_i$ for $1\leq i\leq n$. Let $d: \mathcal{Z} \times \hat{\mathcal{Z}} \to [0,\infty)$ be a distortion measure. Assume $\mathcal{Z} \subseteq \hat{\mathcal{Z}}$ and $d(z,z) = 0, \forall z \in \mathcal{Z}$.

An $(n,2^{nR},t)$ code is defined by an encoding function 
\[
\mathrm{Enc}: \mathcal{S}_1^n \to \{0,1\}^{nR},
\]
a series of subsets $(I_i)_{i =1}^{n}$ such that
\[
|I_i| \leq t, I_i \subseteq \{1,2,...,nR\}, \forall i \in\{1,...,n\},
\]
and a series of local decoding functions $(\mathrm{Dec}_i)_{i =1}^n$  
\[
\mathrm{Dec}_i:\{0,1\}^{|I_i|} \times \mathcal{S}_2^n \to \hat{\mathcal{Z}}, \forall i \in\{1,...,n\}.
\]
Then the encoded message is $M^{nR} = \mathrm{Enc}(S_1^n)$, and its components with indices in $I_i$ are denoted by $M_{I_i}$. Moreover, the reconstruction for each $Z_i$ is $\hat{Z}_i = \mathrm{Dec}_i(M_{I_i},S_2^n)$.
Similarly, the constant $t$ is the locality. 

A rate-tolerance pair $(R,\epsilon)$ is called {\it  achievable} if there exists an $(n,2^{nR},t)$ code such that 
\begin{equation}
\label{eq:achivable}
    \lim_{n \to \infty} \max_{1 \leq i \leq n} \mathbb{P}[d({Z}_i,\hat{Z}_i)> \epsilon] = 0.
\end{equation}
We define the rate-tolerance function $R_{loc}(\epsilon)$ for local decoding to be the infimum of all the rates such that $(R,\epsilon)$ is achievable.

\subsection{Expander Graph Code}
Consider the problem of efficiently storing a sparse vector. Specifically, for an integer $N$ and $\delta \in (0,1)$, let 
\[
S^N_{\delta} = \{x^N \in \{0,1\}^N: |\{i:x_i = 1\}| \leq N \delta\}.
\]
We need an efficient scheme such that, for each sparse $x^N \in S^N_{\delta}$ and any $i \in \{1,...,N\}$, the question ``Is $x_i = 1$?'' can be correctly answered by probing at most constant number (independent of $N$) of bits. 
The coding scheme has been given by the expander graph code in~\cite{Buhrman2002}, where an error bound is obtained for probing only one of a few specified bits. It is further shown in~\cite{Tatwawadi2018} that if all these specified bits are probed, the decoding message can be lossless. The result is summarized as follows.

Define $(N,\delta,R,t)$ expander graph code by an encoding function
\[
\mathrm{Enc}: S^N_{\delta} \to \{0,1\}^{NR},
\]
a series of subsets $(I_i)_{i =1}^{N}$ such that
\[
|I_i| \leq t, I_i \subseteq \{1,2,...,NR\}, \forall i \in \{ 1,...,N\},
\]
and a series of local decoding functions 
\[
\mathrm{Dec}_i: \{0,1\}^{|I_i|} \to \{0,1\}, \forall i \in \{1,...,N\}.
\]
\begin{lemma}
\label{lem:expander}
For any $ \delta \in (0,\frac{1}{4})$, there exists an $(N,\delta,R,t)$ expander graph code such that
\begin{enumerate}
    \item $x_i = \mathrm{Dec}_i((\mathrm{Enc}(x^N))_{I_i})$, for any $x^N \in S^N_{\delta}$ and $i \in \{1,...,N\}$;
    \item $R = O(\delta \log \frac{1}{\delta})$, $t = O(\log \frac{1}{\delta})$.
\end{enumerate}

\end{lemma}

%% file: 3.tex
\section{The Achievable Rate Region and the Layered Coding Scheme}
\label{sec:achievable}
Let $\tilde{\mathscr{R}}_{loc}(\epsilon)$ be the closure of the set of rate pairs $(R_1,R_2)$ such that 
\begin{equation}
\label{eq:innerbound}
\begin{aligned}
    R_1 > I(X_1,U_1|Q),
    \\
    R_2 > I(X_2;U_2|Q),
\end{aligned}
\end{equation}
for some joint distribution $p(x_1,x_2)p(q)p(u_1|x_1,q)p(u_2|x_2,q)$ with $|\mathcal{Q}| \leq 2$, $|\mathcal{U}_1| \leq |\mathcal{X}_1|+2$ and $|\mathcal{U}_2| \leq |\mathcal{X}_2|+2$, and there exists a function $g:\mathcal{U}_1 \times \mathcal{U}_2 \to \hat{Z}$ satisfying $\mathbb{P}[d(Z,g(U_1,U_2)) \leq \epsilon] = 1$.
Then the following theorem shows that $\tilde{\mathscr{R}}_{loc}(\epsilon)$ is achievable.

\begin{theorem}
\label{thm:distributedinner}
$\tilde{\mathscr{R}}_{loc}(\epsilon) \subseteq \mathscr{R}_{loc}(\epsilon)$.
\end{theorem}

\begin{remark}
Consider the distributed lossless compression problem with constant locality. In~\cite{Vatedka2022}, $p(x_1,x_2)$ is called $\mathcal{X}_1$-confusable if, for any $x_2,x_2' \in \mathcal{X}_2$, there exists some $x_1 \in \mathcal{X}_1$ such that $p(x_1,x_2)>0$ and $p(x_1,x_2')>0$. The notion of $\mathcal{X}_2$-confusable is defined similarly. If $p(x_1,x_2)$ is both $\mathcal{X}_1$-confusable and $\mathcal{X}_2$-confusable, the rate region is shown to be $\mathscr{R}_0 \triangleq \{(R_1,R_2)|R_1 \geq H(X_1),R_2 \geq H(X_2)\}$ in~\cite{Vatedka2022}. If $p(x_1,x_2)$ is either not $\mathcal{X}_1$-confusable or not $\mathcal{X}_2$-confusable, Theorem 3 in~\cite{Vatedka2022} shows that the rate region is strictly larger than $\mathscr{R}_0$. However, a general achievable region was not given therein. For this case, specializing \Cref{thm:distributedinner} by letting $f(X_1,X_2) = (X_1,X_2)$ and $\epsilon = 0$, such a general achievable bound can be obtained.
\end{remark}

We prove~\Cref{thm:distributedinner} by designing a layered coding scheme in~\Cref{subsec:coding} and showing its achievability and constant locality property in~\Cref{subsec:analysis}.

Let $(U_1,U_2,g)$ satisfy the Markov chain $U_1-X_1-X_2-U_2$ and $\mathbb{P}[d(Z,g(U_1,U_2)) \leq \epsilon] = 1$. It suffices to show that the rate-tolerance triple $(R_1,R_2,\epsilon)$ satisfying $R_1 > I(X_1;U_1)$ and $R_2 > I(X_2;U_2)$ is achievable, then the bound~\eqref{eq:innerbound} is achieved by time sharing (note that $|\mathcal{Q}|$ can be chosen to be finite). 
We show the achievability by defining a series of $(nb,2^{nbR_1},2^{nbR_2},t)$ distributed code, where $b$ and $t$ are chosen in the following detailed scheme.

\subsection{The Layered Coding Scheme}
\label{subsec:coding}
We use the following definition of robust typicality.
For any $\delta>0$, the set of all {\it $\delta$-robustly typical}~\cite{Orlitsky2001} sequences of length $n$ for a random variable $X$ is defined to be 
\begin{equation}
\label{eq:robusttypical}
    \mathcal{T}^{n,\delta}_{X} = \left\{x^n : \left|\frac{|\{i:x_i = x\}|}{n}- p(x)\right|\leq \delta \cdot p(x), \forall x \in \mathcal{X}\right\}.
\end{equation}

\subsubsection{Codebook Generation}
Our coding scheme consists of three layers, for each of which we design the codebook as follows.

For $k = 1,2$, let $R_{k,1} = I(X_k;U_k) + \delta'$. Then by the proof of the rate-distortion theorem (c.f.~\cite{EIGamal2011}), there exists some $b \in \mathbb{N}$ such that for any $k = 1,2$, there exist a code $\mathcal{C}_{k,1}$  which consists of
an encoding function 
\[
h_{k,1,e}: \mathcal{X}_k^b \to \{0,1\}^{bR_{k,1}},
\]
and a decoding function
\[
h_{k,1,d}:\{0,1\}^{bR_{k,1}} \to \mathcal{U}_   k^b,
\]
satisfying $\mathbb{P}\left[(h_{k,1,d}(h_{k,1,e}(X_{k}^b)),X_{k}^b) \notin \mathcal{T}^{b,\delta''}_{(U_k,X_k)}\right] \leq \delta$ for some $\delta''>0$.

Let $R_{k,2} = \log |\mathcal{X}_k|+1$, $k = 1,2$,  then there is a simple code $\mathcal{C}_{k,2}$ consisting of an encoding function 
\[
h_{k,2,e}: \mathcal{X}_k^b \to \{0,1\}^{bR_{k,2}},
\]
and a decoding function
\[
h_{k,2,d}:\{0,1\}^{bR_{k,2}} \to \mathcal{X}_k^b,
\]
such that for any $x_k^b \in \mathcal{X}_k^b$, $h_{k,2,e}(x_k^b) \neq 0^{b R_{k,2}}$ and $h_{k.2,d}(h_{k,2,e}(x_k^b)) = x_k^b$.

For any $n \in \mathbb{N}$ and $k = 1,2$, there exists a $(nbR_{k,2}, 2\delta, R_{k,3} = O(\delta \log \frac{1}{\delta}),t = O(\log \frac{1}{\delta}))$ expander graph code $\mathcal{C}_{k,3}$ satisfying the properties in~\Cref{lem:expander}. 
Let $R_k = R_{k,1}+R_{k,2}R_{k,3}$. Assume that the $k$-th encoder, $k = 1,2$, knows the decoding function $h_{k,1,d}$, which determines the successive encoding process. 

\subsubsection{Encoding}
\begin{figure}[!t]
\centering
\begin{tikzpicture}
\node at (-5.5,1.75) {\small $X_{k}^{nb}$};
\draw[dashed] (-4.9,1.35) rectangle (3.7,2.15);
\draw (-4.8,1.5) rectangle (-3.6,2);
\draw (-3.6,1.5) rectangle (-2.4,2);
\draw (-2.4,1.5) rectangle (-1.2,2);
\draw (-1.2,1.5) rectangle (0,2);
\draw (0,1.5) rectangle (1.2,2);
\draw (1.2,1.5) rectangle (2.4,2);
\draw (2.4,1.5) rectangle (3.6,2);
\node at (-4.2,1.75) {$X_{k,1}^b$};
\node at (-3,1.75) {$X_{k,2}^b$};
\node at (-1.8,1.75) {$X_{k,3}^b$};
\node at (-0.6,1.75) {...};
\node at (0.6,1.75) {...};
\node at (1.8,1.75) {...};
\node at (3,1.75) {$X_{k,n}^b$};

\draw[->,>=stealth] (-0.6,1.35)--(-0.6,0.85)--(-4.2,0.85)--(-4.2,0.65);
\draw[->,>=stealth] (-0.6,0.85)--(2.25,0.85)--(2.25,-1.2)--(0.75,-1.2)--(0.75,-1.5);
\draw[->,>=stealth] (2.25,-1.2)--(3.75,-1.2)--(3.75,-1.5);
\node at (-2.2,1.05) {$h_{k,1,e}$};
\node at (0.6,1.05) {$h_{k,2,e}$};

\node at (-7.7,0.25) {\small $M_{k,1}^{nbR_{k,1}}$};
\draw[dashed] (-7.1,-0.15) rectangle (-1.3,0.65);
\draw[fill = green!20] (-7,0) rectangle (-6.2,0.5);
\draw[fill = green!20] (-6.2,0) rectangle (-5.4,0.5);
\draw[fill = red!20] (-5.4,0) rectangle (-4.6,0.5);
\draw[fill = green!20] (-4.6,0) rectangle (-3.8,0.5);
\draw[fill = red!20] (-3.8,0) rectangle (-3,0.5);
\draw[fill = green!20] (-3,0) rectangle (-2.2,0.5);
\draw[fill = green!20] (-2.2,0) rectangle (-1.4,0.5);

\draw[->,>=stealth] (-4.2,-0.15)--(-4.2,-0.8)--(2.25,-0.8);
\draw[->,>=stealth] (-4.2,-0.8)--(-4.2,-3.35);
\node at (-5.5,-2.2) {Copy and paste.};
\node at (-1,-0.55) {Decode for $U_{k}^{nb}$ and obtain $\mathcal{J}_k$.};

\node at (-3.35,-1.3) {\small $M_{k,2}^{nbR_{k,2}}$};
\draw[dashed] (-3.1,-2.15) rectangle (7.6,-1.35);
\draw[fill = green!20] (-3,-2) rectangle (-1.5,-1.5);
\draw[fill = green!20] (-1.5,-2) rectangle (0,-1.5);
\draw[fill = red!20] (0,-2) rectangle (1.5,-1.5);
\draw[fill = green!20] (1.5,-2) rectangle (3,-1.5);
\draw[fill = red!20] (3,-2) rectangle (4.5,-1.5);
\draw[fill = green!20] (4.5,-2) rectangle (6,-1.5);
\draw[fill = green!20] (6,-2) rectangle (7.5,-1.5);
\node at (-2.25,-1.75) {\tiny $0^{bR_{k,2}}$};
\node at (-0.75,-1.75) {\tiny $0^{bR_{k,2}}$};
\node at (0.75,-1.75) {\tiny $h_{k,2,e}(X_{k,3}^b)$};
\node at (2.25,-1.75) {...};
\node at (3.75,-1.75) {...};
\node at (5.25,-1.75) {...};
\node at (6.75,-1.75) {\tiny $0^{bR_{k,2}}$};

\node at (-7.7,-3.75) {\small $M_{k}^{nbR_{k}}$};
\draw[dashed] (-7.1,-4.15) rectangle (-1.4,-3.35);
\draw[fill = green!20] (-7,-4) rectangle (-6.2,-3.5);
\draw[fill = green!20] (-6.2,-4) rectangle (-5.4,-3.5);
\draw[fill = red!20] (-5.4,-4) rectangle (-4.6,-3.5);
\draw[fill = green!20] (-4.6,-4) rectangle (-3.8,-3.5);
\draw[fill = red!20] (-3.8,-4) rectangle (-3,-3.5);
\draw[fill = green!20] (-3,-4) rectangle (-2.2,-3.5);
\draw[fill = green!20] (-2.2,-4) rectangle (-1.4,-3.5);
\draw[fill = blue!20] (-1.4,-4) rectangle (0,-3.5);

\node at (0.6,-2.65) {\small $M_{k,3}^{nbR_{k,2}R_{k,3}}$};
\draw[->,>=stealth] (2.25,-2.15)--(2.25,-2.5);
\draw[fill = blue!20] (1.55,-3) rectangle (2.95,-2.5);
\draw[->,>=stealth] (2.25,-3)--(2.25,-3.25)--(-0.7,-3.25)--(-0.7,-3.5);

\end{tikzpicture}
\caption{The layered encoding process for the $k$-th encoder, where the $i$-th blocks in $M_{k,1}^{nbR_{k,1}}$ and $M_{k,2}^{nbR_{k,2}}$ are colored green if $i \in \mathcal{J}_{k}$ and colored red if $i \notin \mathcal{J}_{k}$.}
\label{fig:coding}
\vspace{-0.2cm}
\end{figure}
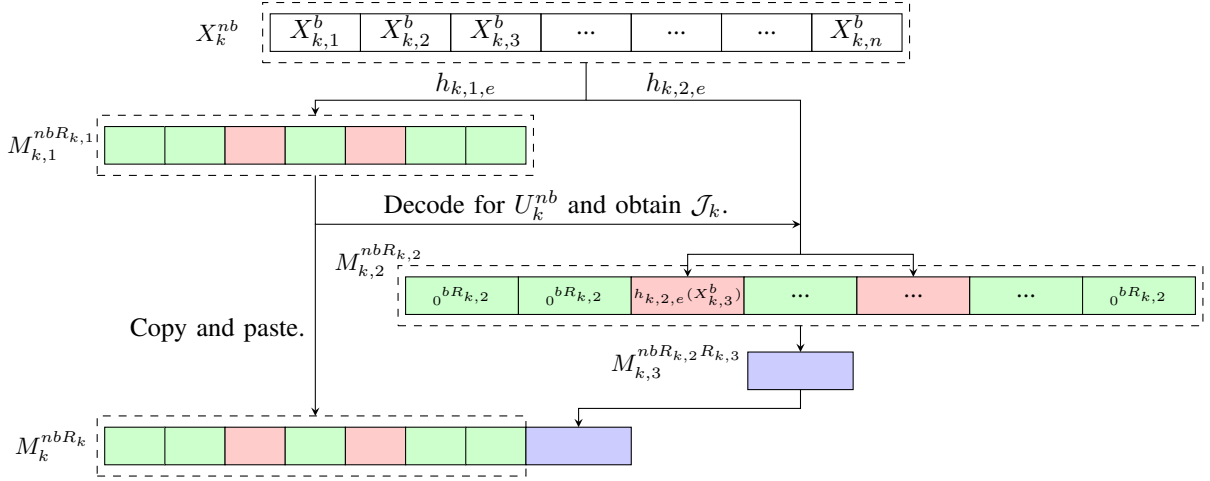

At the $k$-th encoder side, $k = 1,2$, given $X_{k}^{nb} = \big(X_{k,i}^b\big)_{i = 1}^n$, first compute $ M^{nbR_{k,1}}_{k,1} = \big(M^{bR_{k,1}}_{k,1,i}\big)_{i = 1}^n= \big(h_{k,1,e}(X_{k,i}^b)\big)_{i = 1}^n$. Note that the encoder has the luxury to decode the message $M^{nbR_{k,1}}_{k,1}$ generated by itself. That is, it can compute $U_k^{nb} = \big(U_{k,i}^b\big)_{i = 1}^n = \big(h_{k,1,d}(M^{bR_{k,1}}_{k,1,i})\big)_{i = 1}^n$, and $\mathcal{J}_k = \left\{i:(U_{k,i}^b,X_{k,i}^b) \notin \mathcal{T}^{b,\delta''}_{(U_k,X_k)} \right\}$.
Then let $M^{nbR_{k,2}}_{k,2} = \big(M^{bR_{k,2}}_{k,2,i}\big)_{i = 1}^n$, where 
\begin{equation}
    M^{bR_{k,2}}_{k,2,i} = \left\{
    \begin{array}{cc}
         h_{k,2,e}(X_{k,i}^b)& i \in \mathcal{J}_k.  
         \\
         0^{bR_{k,2}}& i \notin \mathcal{J}_k.  
    \end{array}\right.
\end{equation}
After that, it encodes $M^{nbR_{k,2}}_{k,2}$ by the coding scheme of $\mathcal{C}_{k,3}$ into $M^{nbR_{k,2}R_{k,3}}_{k,3}$. 
Finally, let the encoded message be $M^{nbR_k}_k = \big(M^{nbR_{k,1}}_{k,1},M^{nbR_{k,2}R_{k,3}}_{k,3}\big)$, and send it to the decoder. The layered encoding process for the $k$-th encoder is depicted in Fig.~\ref{fig:coding}.

\subsubsection{Decoding}
The decoding process needs partial reconstruction functions $g_1,g_2$ induced by the original reconstruction function $g$ in the following lemma, which is proved in Appendix~\ref{pflem:partial}.

\begin{lemma}
\label{lem:partial}
If $U_1-X_1-X_2-U_2$ is a Markov chain and $\mathbb{P}[d(f(X_1,X_2),g(U_1,U_2)) \leq \epsilon] = 1$, then there exist functions $g_1:\mathcal{U}_1 \times \mathcal{X}_2 \to \hat{Z}$ and $g_2:\mathcal{X}_1 \times \mathcal{U}_2 \to \hat{Z}$ such that $\mathbb{P}[d(f(X_1,X_2),g_1(U_1,X_2)) \leq \epsilon] = \mathbb{P}[d(f(X_1,X_2),g_2(X_1,U_2)) \leq \epsilon] = 1$.
\end{lemma}


Then we describe the decoding process as follows. Given $i \in \{1,...,n\}$ and $j \in \{1,...,b\}$, the decoder constructs $\hat{Z}_{i,j}$ as follows.
First, decode $M^{bR_{k,2}}_{k,2,i}$ for each $k = 1,2$ by probing the specified bits in $M^{nbR_{k,2}R_{k,3}}_{k,3}$. Then decode $\hat{Z}_{i,j}$ according to the values of  $M^{bR_{1,2}}_{1,2,i}$ and $M^{bR_{2,2}}_{2,2,i}$:
\begin{enumerate}[(i)]
\item if $M^{bR_{1,2}}_{1,2,i} = 0^{bR_{1,2}}$ and $M^{bR_{2,2}}_{2,2,i} = 0^{bR_{2,2}}$, then 
compute $U_{1,i}^b = h_{1,1,d}(M^{bR_{1,1}}_{1,1,i})$, $U_{2,i}^b = h_{2,1,d}(M^{bR_{2,1}}_{2,1,i})$ and let $\hat{Z}_{i,j} = g(U_{1,i,j},U_{2,i,j})$;
\item  if $M^{bR_{1,2}}_{1,2,i} = 0^{bR_{1,2}}$ and $M^{bR_{2,2}}_{2,2,i} \neq 0^{bR_{2,2}}$, then 
compute $U_{1,i}^b = h_{1,1,d}(M^{bR_{1,1}}_{1,1,i})$, $X_{2,i}^b = h_{2,2,d}(M^{bR_{2,2}}_{2,2,i})$ and let $\hat{Z}_{i,j} = g_1(U_{1,i,j},X_{2,i,j})$;
\item  if $M^{bR_{1,2}}_{1,2,i} \neq 0^{bR_{1,2}}$ and $M^{bR_{2,2}}_{2,2,i} = 0^{bR_{2,2}}$, then 
compute $X_{1,i}^b = h_{1,2,d}(M^{bR_{1,2}}_{1,2,i})$, $U_{2,i}^b = h_{2,1,d}(M^{bR_{2,1}}_{2,1,i})$ and let $\hat{Z}_{i,j} = g_2(X_{1,i,j},U_{2,i,j})$;
\item  if $M^{bR_{1,2}}_{1,2,i} \neq 0^{bR_{1,2}}$ and $M^{bR_{2,2}}_{2,2,i} \neq  0^{bR_{2,2}}$, then compute $X_{1,i}^b = h_{1,2,d}(M^{bR_{1,2}}_{1,2,i})$, $X_{2,i}^b = h_{2,2,d}(M^{bR_{2,2}}_{2,2,i})$ and let $\hat{Z}_{i,j} = f(X_{1,i,j},X_{2,i,j})$.
\end{enumerate}

\begin{remark}
The first layer codes $\mathcal{C}_{k,1}, k = 1,2$ can be replaced by any  single source rate-distortion codes, 
such as the low density generator matrix (LDGM) code which has lower encoding complexity.

\end{remark}

\subsection{Analysis}
\label{subsec:analysis}

\subsubsection{Code Length Analysis}
The rate of $M^{nbR_k}_k$ is 
\[
R_k = R_{k,1}+R_{k,2}R_{k,3}= I(X_k;U_k)+ \delta'+O\left(\delta \log \frac{1}{\delta}\right),
\]
which can be arbitrarily close to $I(X_k;U_k)$ by letting $\delta'$ and $\delta$ be small enough.

\subsubsection{Number of Probed Bits}
For $k = 1,2$, the number of probed bits in $M^{nbR_k}_k$ for decoding $\hat{Z}_{i,j}$ is bounded by
\[
bR_{k,2}t + bR_{k,1} = b\left[O\left(\log \frac{1}{\delta}\right)+I(U_k;X_k)+ \delta'\right],
\]
which is a constant independent of $n$.

\subsubsection{Error Analysis}
Finally, we prove~\eqref{eq:distrachivable} holds as follows.
For $k = 1,2$, let 
\begin{equation}
    E_{k,i} = \mathds{1}\big\{(U_{k,i}^b,X_{k,i}^b) \notin \mathcal{T}^{b,\delta''}_{(U_k,X_k)} \big\}, i \in \{1,...,n\}.
\end{equation}
Then $E_{k,i}$ are Bernoulli i.i.d. random variables satisfying $\mathbb{P}[E_{k,i} = 1] \leq \delta$. We have $\mathcal{J}_k = \{i: E_{k,i} = 1\}$ and we can use the following inequality to bound the probability that $|\mathcal{J}_k| > 2n\delta$.

\begin{lemma}[Chernoff Bound]
\label{lem:Chernoff}
Let $B_i \in \mathrm{Bern}(p), i\in\{1,2,\cdots,n\}$ be i.i.d. random variables, then for any $t>0$,
\begin{equation}
    \mathbb{P}\left[\sum_{i =1}^n B_i> (1+t)np\right] \leq \exp\left(-\frac{npt^2}{t+2}\right).
\end{equation}
\end{lemma}

Since $|\mathcal{J}_k| = \sum_{i = 1}^nE_{k,i}$,  by~\Cref{lem:Chernoff}, we have
\begin{equation}
\label{eq:E_k}
    \mathbb{P}\left[|\mathcal{J}_k|> 2n\delta\right] \leq e^{-\frac{n\delta}{3}}.
\end{equation}
Let $\mathcal{E}_k$ be the event that $|\mathcal{J}_k| > 2n\delta$, $k = 1,2$. Then for the event $\mathcal{E}_1^c \cap \mathcal{E}_2^c$, we have $|\mathcal{J}_k| \leq 2n \delta$, $k = 1,2$, and by~\Cref{lem:expander} the docoding result of $M^{bR_{k,2}}_{k,2,i}$ is always correct for all $i \in \{1,...,n\}$. In this case, we further show the final reconstruction~$\hat{Z}_{i,j}$ satisfies $\mathbb{P}[d(Z_{i,j},\hat{Z}_{i,j}) \leq \epsilon|\mathcal{E}_1^c \cap \mathcal{E}_2^c] = 1$ for all $i \in \{1,...,n\}$, $j \in\{1,...,b\}$, as follows.

\begin{enumerate}[(a)]
\item 
For the event $\mathcal{E}_1^c \cap \mathcal{E}_2^c  \cap \{i \notin \mathcal{J}_1\} \cap  \{i \notin \mathcal{J}_2\}$, we have $(U_{k,i}^b,X_{k,i}^b) \in \mathcal{T}^{b,\delta''}_{(U_k,X_k)}$, $k = 1,2$.
Then by~\eqref{eq:robusttypical} we have $p(u_k,x_k)>0$, $k = 1,2$, for $(U_{k,i,j},X_{k,i,j}) = (u_k,x_k)$. 
Hence $p(x_1,x_2,u_1,u_2) = p(x_1,x_2)p(u_1|x_1)p(u_2|x_2)>0$. Since $(U_1,U_2,g)$ satisfies $\mathbb{P}[d(f(X_1,X_2),g(U_1,U_2))\leq \epsilon]=1$, we have $d(f(x_1,x_2),g(u_1,u_2)) \leq \epsilon$. By case (i) in the decoding process,  $\hat{Z}_{i,j} = g(U_{1,i,j},U_{2,i,j})$, thus we have $d(Z_{i,j}, \hat{Z}_{i,j})\leq \epsilon$.
\item 
For the event $\mathcal{E}_1^c \cap \mathcal{E}_2^c \cap \{i \in \mathcal{J}_1\} \cap \{i \notin \mathcal{J}_2\}$ or $\mathcal{E}_1^c \cap \mathcal{E}_2^c \cap \{i \notin \mathcal{J}_1\} \cap \{i \in \mathcal{J}_2\}$, similarly by case (ii) and (iii) in the decoding process, we have $d(Z_{i,j}, \hat{Z}_{i,j})\leq \epsilon$.
\item
For the event $\mathcal{E}_1^c \cap \mathcal{E}_2^c \cap \{i \in \mathcal{J}_1\} \cap \{i \in \mathcal{J}_2\}$, by case (iv) in the decoding process we have $d(Z_{i,j}, \hat{Z}_{i,j})= 0$. 
\end{enumerate}
Combining all these cases, by~\eqref{eq:E_k} and the union bound, we obtain
\[
\mathbb{P}[d(Z_{i,j}, \hat{Z}_{i,j})> \epsilon] \leq \mathbb{P}[\mathcal{E}_1]+\mathbb{P}[\mathcal{E}_2] \to 0, n \to \infty,
\]
which completes the proof.

\begin{remark}
\label{rem:block}
If it is the whole block $\hat{Z}^{nb}$ instead of a single $\hat{Z}_{i,j}$ that is constructed by the local decoding scheme, then the above argument implies a vanishing block error probability, i.e., $\mathbb{P}[\exists i,j, d(Z_{i,j},\hat{Z}_{i,j}) > \epsilon] \leq \mathbb{P}[\mathcal{E}_1]+\mathbb{P}[\mathcal{E}_2] \to 0$. 
\end{remark}

%% file: 4.tex
\section{The Converse}
\label{sec:converse}

\subsection{The Converse Theorem}

It is immediate that $\mathscr{R}_{loc}(\epsilon)$ (c.f. ~\Cref{subsec:distributed}) is subsumed in the rate region of 
the classical problem without the locality constraint. Hence the outer bound for the classical problem is applicable for the problem here. For example, an outer bound can be easily obtained by slightly modifying the Berger-Tung outer bound~\cite{Berger1978,Tung1978}.  
Moreover, the following intuitive outer bound can be established.
\begin{lemma}
\label{lem:distributedouter}
An outer bound for $\mathscr{R}_{loc}(\epsilon)$ is characterized by 
\begin{equation}
\begin{aligned}
    R_1 \geq & I(X_1,U_1),
    \\
    R_2 \geq & I(X_2;U_2),
    \\
    R_1 + R_2 \geq & I(X_1,X_2;U_1,U_2),
\end{aligned}
\end{equation}
where $p(u_1,u_2,|x_1,x_2)$ satisfies Markov chains $U_1-X_1-X_2$ and $X_1-X_2-U_2$, and there exists a function $g:\mathcal{U}_1 \times \mathcal{U}_2 \to \hat{Z}$ such that $\mathbb{P}[d(Z,g(U_1,U_2)) \leq \epsilon] = 1$.
\end{lemma}

\begin{IEEEproof}
    The proof can be found in Appendix~\ref{pflem:distributedouter}.
\end{IEEEproof}

Note that the main differences between the inner bound $\tilde{\mathscr{R}}_{loc}(\epsilon)$ and the above outer bound are the Markov constraints. The Markov chain
$U_1-X_1-X_2-U_2$ holds for extreme points of the region $\tilde{\mathscr{R}}_{loc}(\epsilon)$. However, in the outer bound by~\Cref{lem:distributedouter}, two separate pieces of the Markov chain are imposed. 
\Cref{lem:distributedouter} does not exploit the local decoding property, hence it is generally not tight. 

The following theorem shows that under mild regularity conditions, $\tilde{\mathscr{R}}_{loc}(\epsilon)$ is tight. The proof is given in~\Cref{pfthm:distributedexact}.

\begin{theorem}[]
\label{thm:distributedexact}
If $(X_1,X_2)$ has a full support, i.e., $p(x_1,x_2)>0$ for all $x_1 \in \mathcal{X}_1$ and $x_2 \in \mathcal{X}_2$, then $\mathscr{R}_{loc}(\epsilon) = \tilde{\mathscr{R}}_{loc}(\epsilon)$. 
\end{theorem}

\begin{remark}
Note that for the classical problem without the constant locality constraint, an achievable inner bound $\mathscr{R}_{BT}(\epsilon)$ can be established by the Berger-Tung coding~\cite{Berger1978,Tung1978}, where $\mathscr{R}_{BT}(\epsilon)$ is the closure of the set of rate pairs $(R_1,R_2)$ such that 
\begin{align*}
    R_1 &> I(X_1,U_1|U_2,Q),
    \\
    R_2 &> I(X_2;U_2|U_1,Q),
    \\
    R_1+R_2 &> I(X_1,X_2;U_1,U_2|Q)
\end{align*}
for some joint distribution $p(x_1,x_2)p(q)p(u_1|x_1,q)p(u_2|x_2,q)$, and there exists a function $g:\mathcal{U}_1 \times \mathcal{U}_2 \to \hat{Z}$ satisfying $\mathbb{P}[d(Z,g(U_1,U_2)) \leq \epsilon] = 1$.
It is easy to verify that $\tilde{\mathscr{R}}_{loc}(\epsilon) \subseteq \mathscr{R}_{BT}(\epsilon)$ and the inclusion is generally strict. This together with ~\Cref{thm:distributedexact} implies that in most cases the constant locality constraint force the rate region to shrink.  Therefore a larger rate is necessary in order to satisfy the locality constraint.
\end{remark}

Next, we specialize \Cref{thm:distributedexact} to the following corollaries. First let $\mathcal{Z} = \hat{\mathcal{Z}}$, $d(z,\hat{z}) = \mathds{1}\{z \neq \hat{z}\}$ and $\epsilon = 0$, then the rate region for distributed lossless computing is easily obtained as follows. 

\begin{corollary}
If $(X_1,X_2)$ has a full support, then the rate region for the distributed lossless computing problem with constant locality is characterized by
\begin{equation}
\begin{aligned}
    R_1 \geq I(X_1,U_1|Q),
    \\
    R_2 \geq I(X_2;U_2|Q),
\end{aligned}
\end{equation}
for some joint distribution $p(x_1,x_2)p(q)p(u_1|x_1,q)p(u_2|x_2,q)$ with $|\mathcal{Q}| \leq 2$, $|\mathcal{U}_1| \leq |\mathcal{X}_1|+2$ and $|\mathcal{U}_2| \leq |\mathcal{X}_2|+2$, and there exists a function $g:\mathcal{U}_1 \times \mathcal{U}_2 \to \hat{Z}$ such that $Z = g(U_1,U_2)$ almost surely.
\end{corollary}

Then consider the distributed approximate compression problem, i.e., $f(x_1,x_2) = (x_1,x_2)$ and $d((x_1,x_2)),(\hat{x}_1,\hat{x}_2)) = \max\{d_1(x_1,\hat{x}_1),d_2(x_2,\hat{x}_2)\}$. The rate region is also implied from~\Cref{thm:distributedexact}.

\begin{corollary}
\label{cor:dac}
If $(X_1,X_2)$ has a full support, then the rate region for the distributed approximate compression problem is characterized by
\begin{equation}
\label{eq:dacbound}
\begin{aligned}
    R_1 \geq \min_{\substack{p(u_1|x_1): 
    \exists g_1 \\ \mathbb{P}[d_1(X_1,g_1(U_1)) \leq \epsilon] = 1}} I(X_1,U_1), \\
    R_2 \geq \min_{\substack{p(u_2|x_2): 
    \exists g_2 \\ \mathbb{P}[d_2(X_2,g_2(U_2)) \leq \epsilon] = 1}} I(X_2,U_2),
\end{aligned}
\end{equation}
with $|\mathcal{U}_1| \leq |\mathcal{X}_1|+1$ and $|\mathcal{U}_2| \leq |\mathcal{X}_2|+1$.
\end{corollary}

\begin{IEEEproof}
    The proof can be found in Appendix~\ref{pfcor:dac}.
\end{IEEEproof}

The rate region given by \Cref{cor:dac} can be alternatively achieved by separate coding at two sides. It is intuitive that separate compression is optimal, since the locality constraints require the encoders can access to only a relative small constant number of bits, which do not reflect the correlation of sources.

\subsection{Proof of Theorem~\ref{thm:distributedexact}}
\label{pfthm:distributedexact}
We now give the proof of~\Cref{thm:distributedexact}.
The idea is to consider the local decoding process as a inference problem from $(M_{1,I_{1,i}},M_{2,I_{2,i}})$ to $Z_i$.
By the reverse hypercontractivity property, each encoder side has to give enough information for constructing $M^{nR_k}_k$, so that the distribution of $X_{k,i}$ that is useful for decoding $Z_i$ can be inferred from $M_{k,I_{k,i}}$.
Then $p_{M_{k,I_{k,i}}|X_{k,i}}$ resembles $p_{U_{k,i}|X_{k,i}}$ for some $U_{k,i}$ satisfying the constraint in~\eqref{eq:innerbound} for the auxiliary random variable. Hence $U_{k,i}$ can be constructed by shifting a small amount of total probability based on the distribution of $M_{k,I_{k,i}}$, and $I(X_{k,i};M_{k,I_{k,i}})$ can be lower bounded by $ I(X_{k,i};U_{k,i})$ using Fano's inequality. The bound for $R_k$ is then obtained.

Consider an $(n,2^{nR_1},2^{nR_2},t)$ code achieving the rate-tolerance triple $(R_1,R_2,\epsilon)$. Then for any $\delta>0$,  there exists some $N \in \mathbb{N}$, such that for any $n>N$ and $i \in \{1,...,n\}$ we have $\mathbb{P}[d(Z_i,\hat{Z}_i)> \epsilon] \leq \delta$.
Hence for any $m_1 \in \{0,1\}^{|I_{1,i}|}$ and $m_2 \in \{0,1\}^{|I_{2,i}|}$, we have
\begin{align*}
         \delta \geq& \mathbb{P}[d(Z_i,\hat{Z}_i)> \epsilon]
         \\
         \geq&\mathbb{P}[d(Z_i,\hat{Z}_i)> \epsilon, M_{1,I_{1,i}} = m_1,M_{2,I_{2,i}} = m_2] 
         \\
         \geq & \min_{\hat{z} \in \mathcal{\hat{Z}}} \mathbb{P}[d(Z_i, \hat{z}) >\epsilon,M_{1,I_{1,i}} = m_1,M_{2,I_{2,i}} = m_2] 
         \\
         = & \min_{\hat{z} \in \mathcal{\hat{Z}}} \sum_{\substack{x_1,x_2: \\d(f(x_1,x_2),\hat{z})> \epsilon}} \mathbb{P}[X_{1,i} = x_1,M_{1,I_{1,i}} = m_1,X_{2,i} = x_2,M_{2,I_{2,i}} = m_2].
\end{align*}

We introduce the following useful lemma (c.f. Lemma 8.3 in~\cite{Mossel2013}) to bound the joint probability appeared above by its marginal. 

\begin{lemma}[Reverse Hypercontractivity]
\label{lem:rehy}
Let $(V_i,W_i) \sim p(v,w), i = \{1,...,n\}$ be i.i.d. random variables distributed over a finite alphabet $\mathcal{V} \times \mathcal{W}$. Suppose $p(v,w)$ has a full support, i.e., $p(v,w)>0$ for any $v \in \mathcal{V}$ and $w \in \mathcal{W}$. Then there exists $\alpha, \beta \in (1,\infty)$ depending only on $p(v,w)$, such that for any subsets $\mathcal{A} \subseteq \mathcal{V}^n$ and $\mathcal{B} \subseteq \mathcal{W}^n$, we have
\begin{equation}
    \mathbb{P}[V^n \in \mathcal{A}, W^n \in \mathcal{B}] \geq (\mathbb{P}[V^n \in \mathcal{A}])^{\alpha}(\mathbb{P}[W^n \in \mathcal{B}])^{\beta}.
\end{equation}
\end{lemma}

Since $(X_1,X_2)$ has a full support, by \Cref{lem:rehy}, we have 
\begin{equation}
\label{eq:rever}
\begin{aligned}
    &\mathbb{P}[X_{1,i} = x_1,M_{1,I_{1,i}} = m_1,X_{2,i} = x_2,M_{2,I_{2,i}} = m_2] 
    \\
    \geq &(\mathbb{P}[X_{1,i} = x_1,M_{1,I_{1,i}} = m_1])^{\alpha_1} (\mathbb{P}[X_{2,i} = x_2,M_{2,I_{2,i}} = m_2])^{\alpha_2},
\end{aligned}
\end{equation}
where $\alpha_1, \alpha_2 >1$ only depend on $p(x_1,x_2)$.

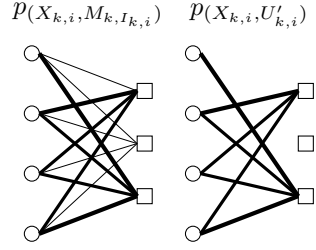
\begin{figure}[!t]
\centering
\begin{tikzpicture}
\node at (2.75,0.5) {$p_{(X_{k,i},M_{k,I_{k,i}})}$};

\draw (2,0) circle(0.1);

\draw (2,-0.8) circle(0.1);

\draw (2,-1.6) circle(0.1);

\draw (2,-2.4) circle(0.1);

\draw (3.4,-0.6) rectangle (3.6,-0.4);

\draw (3.4,-1.3) rectangle (3.6,-1.1);

\draw (3.4,-2) rectangle (3.6,-1.8);

\draw[ultra thin]  (2.1,0)--(3.4,-0.5);
\draw[ultra thin]  (2.1,0)--(3.4,-1.2);
\draw[ultra thick]  (2.1,0)--(3.4,-1.9);

\draw[ultra thick]  (2.1,-0.8)--(3.4,-0.5);
\draw[ultra thin]  (2.1,-0.8)--(3.4,-1.2);
\draw[very thick]  (2.1,-0.8)--(3.4,-1.9);

\draw[very thick]  (2.1,-1.6)--(3.4,-0.5);
\draw[ultra thin]  (2.1,-1.6)--(3.4,-1.2);
\draw[very thick]  (2.1,-1.6)--(3.4,-1.9);

\draw[very thick]  (2.1,-2.4)--(3.4,-0.5);
\draw[ultra thin]  (2.1,-2.4)--(3.4,-1.2);
\draw[ultra thick] (2.1,-2.4)--(3.4,-1.9);
\end{tikzpicture}
\begin{tikzpicture}
\node at (2.75,0.5) {$p_{(X_{k,i},U'_{k,i})}$};

\draw (2,0) circle(0.1);

\draw (2,-0.8) circle(0.1);

\draw (2,-1.6) circle(0.1);

\draw (2,-2.4) circle(0.1);

\draw (3.4,-0.6) rectangle (3.6,-0.4);

\draw (3.4,-1.3) rectangle (3.6,-1.1);

\draw (3.4,-2) rectangle (3.6,-1.8);

\draw[ultra thick]  (2.1,0)--(3.4,-1.9);

\draw[ultra thick]  (2.1,-0.8)--(3.4,-0.5);
\draw[very thick]  (2.1,-0.8)--(3.4,-1.9);

\draw[very thick]  (2.1,-1.6)--(3.4,-0.5);
\draw[very thick]  (2.1,-1.6)--(3.4,-1.9);

\draw[very thick]  (2.1,-2.4)--(3.4,-0.5);
\draw[ultra thick] (2.1,-2.4)--(3.4,-1.9);
\end{tikzpicture}
\caption{Illustration for the construction of $U_{k,i}'$, where the line width represents the probability of the pair $(x_k,m_k)$. For each $x_k \in \mathcal{X}_k$, the sufficiently small joint probabilities to some $m_k \in \{0,1\}^{|I_{k,i}|}$ is shifted to $m_{k,x_k}$ with a large joint probability.}
\label{fig:construction}
\vspace{-0.2cm}
\end{figure}

Let $\mathcal{U}_{k,i} = \{0,1\}^{|I_{k,i}|}$, $k = 1,2$.
If $\delta$ is small enough (i.e., $2^t \cdot \delta^{\frac{1}{1+2\alpha_k}}< \min_{x_k \in \mathcal{X}_k} p(x_k), \forall k = 1,2$), then for any $x_k$, there exists some $m_{k,x_k} \in \{0,1\}^{|I_{k,i}|}$ such that $p_{(X_{k,i},M_{k,I_{k,i}})}(x_k,m_{k,x_k})> \delta^{\frac{1}{1+2\alpha_k}}$, $k = 1,2$. 
Define random variables $U'_{k,i} = h_{k}(X_{k,i},M_{k,I_{k,i}}) \in \mathcal{U}_{k,i}$, $k = 1,2$, where
\begin{equation}
\label{eq:defhk}
    h_{k}(x_k,m_k) = 
    \left\{
    \begin{array}{cc}
         m_k, & \text{if } p_{(X_{k,i},M_{k,I_{k,i}})}(x_k,m_k)> \delta^{\frac{1}{1+2\alpha_k}},
         \\
         m_{k,x_k}, & \text{if }  p_{(X_{k,i},M_{k,I_{k,i}})}(x_k,m_k)\leq \delta^{\frac{1}{1+2\alpha_k}}. 
    \end{array}
    \right.
\end{equation}
In other words, the distribution of $(X_{k,i},U_{k,i}')$ is constructed from that of $(X_{k,i},M_{k,I_{k,i}})$ by shifting probabilities to large probabilities. See~\Cref{fig:construction} for illustration.

Then define random variables $U_{k,i} \in \mathcal{U}_{k,i}$, $k = 1,2$, by letting the joint distribution be 
\begin{align*}
&p(x_{1,i},x_{2,i},u_{1,i},u_{2,i}) 
\\
=&p(x_{1,i},x_{2,i})p_{U_{k,i}|X_{k,i}}(u_{1,i}|x_{1,i})p_{U_{k,i}|X_{k,i}}(u_{2,i}|x_{2,i}),
\end{align*}
where $p_{U_{k,i}|X_{k,i}} = p_{U'_{k,i}|X_{k,i}}$.
Next we show there exists a function $g$ such that $\mathbb{P}[d(f(X_{1,i},X_{2,i}),g(U_{1,i},U_{2,i})) \leq \epsilon] = 1$. It is equivalent to show that, for any $(u_1,u_2)$, there exists some $\hat{z} \in \hat{\mathcal{Z}}$ such that for any $(x_1,x_2)$ with $p(x_1,u_1)>0$ and $p(x_2,u_2)>0$, we have $d(f(x_1,x_2),\hat{z}) \leq \epsilon$. 
Suppose not, then there exists some $(u_1,u_2)$ such that for any $\hat{z} \in \hat{\mathcal{Z}}$, there exists some pair $(x_1,x_2)$ with $p(x_1,u_1)>0$,  $p(x_2,u_2)>0$ and $d(f(x_1,x_2),\hat{z}) > \epsilon$.
For such a pair $(x_1,x_2)$, since $p_{U_{k,i}|X_{k,i}} = p_{U'_{k,i}|X_{k,i}}$, we have 
\begin{align*}
    p(x_k,u_k) = &\mathbb{P}[X_{k,i} = x_k,U_{k,i} = u_k]
    \\
    =&\mathbb{P}[X_{k,i} = x_k,U'_{k,i} = u_k],
\end{align*}
which implies $\mathbb{P}[X_{k,i} = x_k,M_{k,I_{k,i}} = u_k] > \delta^{\frac{1}{1+2\alpha_k}}$, $k = 1,2$ by~\eqref{eq:defhk}.
Then by~\eqref{eq:rever},  
\[
\mathbb{P}[X_{1,i} = x_1,M_{1,I_{1,i}} = m_1,X_{2,i} = x_2,M_{2,I_{2,i}} = m_2] > \delta^{\frac{\alpha_1}{1+2\alpha_1}+\frac{\alpha_2}{1+2\alpha_2}}.
\]
By the existence of $(x_1,x_2)$ for each $\hat{z}$, we have 
\begin{align*}
    \min_{\hat{z} \in \mathcal{\hat{Z}}} \sum_{\substack{x_1,x_2: \\d(f(x_1,x_2),\hat{z})> \epsilon}} \mathbb{P}[X_{1,i} = x_1,M_{1,I_{1,i}} = m_1,X_{2,i} = x_2,M_{2,I_{2,i}} = m_2]>\delta, 
\end{align*}
which leads to a contradiction. Thus we obtain $(U_{1,i},U_{2,i},g)$ satisfying the Markov chain $U_{1,i}-X_{1,i}-X_{2,i}-U_{2,i}$ and $\mathbb{P}[d(Z_i,g(U_{1,i},U_{2,i})) \leq \epsilon] = 1$.

By~\eqref{eq:defhk}, we have $\mathbb{P}[U'_{k,i} \neq M_{k,I_{k,i}}] \leq 2^t \cdot |\mathcal{X}_k| \cdot \delta^{\frac{1}{1+2\alpha_k}}$.
Then by Fano's inequality, we have
\begin{align*}
    &H(U_{k,i}'|M_{k,I_{k,i}}) 
    \\
    \leq &h_{b}(2^t \cdot |\mathcal{X}_k| \cdot \delta^{\frac{1}{1+2\alpha_k}})+2^t \cdot |\mathcal{X}_k| \cdot \delta^{\frac{1}{1+2\alpha_k}} \log  (2^t) 
    \\
    \triangleq &c(k,\delta)
\end{align*}
where $h_b$ is the binary entropy function, and $c(k,\delta) \to 0$ for $ \delta \to 0$.
Since $p_{U_{k,i}|X_{k,i}} = p_{U'_{k,i}|X_{k,i}}$, we have 
\begin{align*}
    &I(X_{k,i};M_{k,I_{k,i}}) - I(X_{k,i};U_{k,i})
    \\
    =&I(X_{k,i};M_{k,I_{k,i}}) - I(X_{k,i};U'_{k,i})
    \\
    =&H(X_{k,i}|U'_{k,i})-H(X_{k,i}|M_{k,I_{k,i}})
    \\
    \geq &H(X_{k,i}|U'_{k,i},M_{k,I_{k,i}})-H(X_{k,i},U'_{k,i}|M_{k,I_{k,i}})
    \\
    = & -H(U'_{k,i}|M_{k,I_{k,i}})
    \\
    \geq &-c(k,\delta).
\end{align*}

Finally, we bound $R_k$ by the following lemma, which is proved in Appendix~\ref{pflem:averagerate}.
\begin{lemma}
\label{lem:averagerate}
For any $(n,2^{nR_1},,2^{nR_2},t)$ distributed code, we have
\begin{equation}
    R_k \geq \frac{1}{n}\sum_{i =1}^n I(X_{k,i};M_{k,I_{k,i}}), k = 1,2.
\end{equation}
\end{lemma}

By~\Cref{lem:averagerate}, for $k = 1,2$, we have
\begin{align*}
    R_k \geq & \frac{1}{n}\sum_{i =1}^n I(X_{k,i};M_{k,I_{k,i}})
    \\
    \geq &\frac{1}{n}\sum_{i =1}^n I(X_{k,i};U_{k,i}) - c(k,\delta)
    \\
    =&I(X_k^{(n)};U_k^{(n)}|Q)- c(k,\delta),
\end{align*}
where the inequality is obtained by letting $Q \sim \mathrm{Unif}[1:n]$ independent of $(X_1^n,X_2^n)$, $X_1^{(n)} = X_{1,Q}$, $X_2^{(n)} = X_{2,Q}$, $U_1^{(n)} = (U_{1,Q})$, $U_2^{(n)} = (U_{2,Q})$ and $\hat{Z}^{(n)} = \hat{Z}_Q$. Letting $\delta \to 0$ completes the proof.

\begin{remark}
\label{rem:averageerr}
Note that the reconstruction quality constraint~\eqref{eq:distrachivable} can be weakened to an average one
\begin{equation}
\label{eq:distrachievable1}
    \lim_{n \to \infty}\frac{1}{n} \sum_{i = 1}^n \mathbb{P}[d({Z}_i,\hat{Z}_i)> \epsilon] = 0,
\end{equation}
and the conclusion of~\Cref{thm:distributedexact} remains true.
To see this, fix $N \in \mathbb{N}$. Then~\eqref{eq:distrachievable1} implies that for any $\delta >0$, as long as $n$ is sufficiently large, $\frac{1}{n} \sum_{i = 1}^n \mathbb{P}[d({Z}_i,\hat{Z}_i)> \epsilon] \leq \frac{\delta}{N}$. Hence for at least $n\left(1- \frac{1}{N}\right)$ many $i \in \{1,...,n\}$, we have $\mathbb{P}[d({Z}_i,\hat{Z}_i)> \epsilon] \leq \delta$. Thus by the proof of~\Cref{thm:distributedexact} we have 
\begin{align*}
    R_k \geq & \left(1-\frac{1}{N}\right) \left(I(X_k;U_k|Q)- c(k,\delta)\right), k = 1,2.
\end{align*}
Letting $\delta \to 0$ and $N \to \infty$ completes the proof.
\end{remark}

\begin{remark}
\label{rem:zerodistortion}
Furthermore, let $\epsilon = 0$. Since both $\mathcal{Z}$ and $\hat{\mathcal{Z}}$ are finite, we have $c d(z,\hat{z}) \leq \mathds{1}\{d(z,\hat{z})> 0\} \leq C d(z,\hat{z})$ for some $C,c>0$. Hence the constraint~\eqref{eq:distrachievable1} can be replaced by
 \begin{equation}
 \label{eq:destrachievable2}
    \lim_{n \to \infty}\frac{1}{n} \sum_{i = 1}^n \mathbb{E}[d({Z}_i,\hat{Z}_i)] = 0,
\end{equation}
and the conclusion of~\Cref{thm:distributedexact} still remains true.
\end{remark}

\begin{remark}
Similar to~\Cref{rem:averageerr}, the worst case decoding locality constraint~\eqref{eq:worstlocality} can be weakened to an average one,
\begin{equation}
\label{eq:averagelocality}
\begin{aligned}
    \frac{1}{n} \sum_{i = 1}^{n} |I_{1,i}| \leq t,
    \\
    \frac{1}{n} \sum_{i = 1}^{n} |I_{2,i}| \leq t,
\end{aligned} 
\end{equation}
without changing the result of~\Cref{thm:distributedexact}. Let $G_n$ be the {\it factor graph}  of the decoding functions $(\mathrm{Dec}_i)_{i = 1}^n$, where each reconstructed symbol $\hat{Z}_i$ is connected to the probed bits in $M_{1,I_{1,i}}$ and $M_{2,I_{2,i}}$ in the graph. Then~\eqref{eq:averagelocality} is equivalent the sparsity of $G_n$. To be precise, sparsity means that the number of edges in $G_n$ increases linearly with $n$, which is proportional to the number of vertices. It is clear that low decoding complexity is reflected by the sparsity of $G_n$, as well as the decoding locality constraint~\eqref{eq:averagelocality}.
\end{remark}

%% file: 5.tex
\section{Generalization to More than Two Encoders}
\label{sec:more}
It is straightforward to generalize the distributed computing problem to the case of $m>2$ sources.
Let $X_k, k \in \{1,...,m\}$ be the distributed sources and $Z= f(X_1,...,X_m)$ be the function to be computed. 
Other settings are similar to that in~\Cref{subsec:distributed}.
Denote the rate region with $m>2$ encoders by $\mathscr{R}^m_{loc}(\epsilon)$. Then the results for two encoders can be easily generalized to the following theorems. We omit the proofs that are similar to that in~\Cref{sec:achievable,sec:converse}.

Let $\tilde{\mathscr{R}}^{m}_{loc}(\epsilon)$ be the closure of the set of rate pairs $(R_1,...,R_m)$ such that
\begin{equation}
\begin{aligned}
    R_k > I(X_k,U_k|Q), \forall k \in \{1,...,m\}.
\end{aligned}
\end{equation}
for some joint distribution $p(x_1,...,x_m)p(q)\prod_{k = 1}^m p(u_k|x_k,q)$ with $|\mathcal{Q}| \leq m$ and $|\mathcal{U}_k| \leq |\mathcal{X}_k|+m, k \in \{1,...,m\}$, and there exists a function $g:\prod_{k = 1}^m \mathcal{U}_k \to \hat{Z}$ such that $\mathbb{P}[d(Z,g(U_1,...,U_m)) \leq \epsilon] = 1$. 

The following theorem shows that $\tilde{\mathscr{R}}^{m}_{loc}(\epsilon)$ is achievable for general sources.

\begin{theorem}
$\mathscr{R}^m_{loc}(\epsilon) \subseteq \tilde{\mathscr{R}}^{m}_{loc}(\epsilon)$.
\end{theorem}

An outer bound for general sources can be similarly established.
\begin{lemma}
An outer bound is characterized by
\begin{equation}
\begin{aligned}
    \sum_{k \in S} R_k \geq & I(X_{\mathcal{S}},U_{\mathcal{S}}), \forall \mathcal{S} \subseteq \{1,...,m\},
\end{aligned}
\end{equation}
where $p(u_1,...,u_m|x_1,...,x_m)$ satisfies $p(u_k|x_1,...,x_m) = p(u_k|x_k), k\in \{1,...,m\}$, and there exists a function $g:\prod_{k = 1}^m \mathcal{U}_k \to \hat{Z}$ such that $\mathbb{P}[d(Z,g(U_1,...,U_m)) \leq \epsilon] = 1$.
\end{lemma}

Moreover, the rate region $\tilde{\mathscr{R}}^{m}_{loc}(\epsilon)$ is tight for sources with a full support.

\begin{theorem}[]
If $p(x_1,...,x_m)>0$ for any $x_k \in \mathcal{X}_k, k\in \{1,...,m\}$, then $\mathscr{R}_{loc}^m(\epsilon)=\tilde{\mathscr{R}}^{m}_{loc}(\epsilon)$.
\end{theorem}

%% file: 6.tex
\section{Constant Locality in Approximate Computing with Side Information}
\label{sec:sideinformation}
In this section, we establish the results for the problem of computing with side information (c.f.~\Cref{subsec:sideinformation}) parallel to that for the distributed computing problem. First let 
\begin{equation}
\label{eq:R(e)}
    \tilde{R}_{loc}(\epsilon) = \min_{\substack{U-S_1-S_2, \\ \exists g: \mathbb{P}[d(Z,g(U,S_2))\leq \epsilon]=1}} I(U;S_1),
\end{equation}
then the following theorem shows that the rate $\tilde{R}_{loc}(\epsilon)$ is achievable.

\begin{theorem}
\label{thm:sideinformationinner}
We have $R_{loc}(\epsilon) \leq \tilde{R}_{loc}(\epsilon)$.
\end{theorem}

\begin{IEEEproof}
The rate $\tilde{R}_{loc}(\epsilon)$ can be achieved by a simplified  layered coding scheme, for which details can be found in Appendix~\ref{pfthm:sideinformationinner}.
\end{IEEEproof}

\begin{remark}
We can add a constraint $|\mathcal{U}| \leq |\mathcal{S}_1|+1$ to the minimum in~\eqref{eq:R(e)} without changing the value of $\tilde{R}_{loc}(\epsilon)$.
\end{remark}

\begin{remark}
Similar to~\Cref{rem:block}, if it is the whole block $\hat{Z}^{nb}$ instead of a single $\hat{Z}_{i,j}$ that is constructed by the constant locality scheme, then the proof shows a vanishing block error probability, i.e., $\mathbb{P}[\exists i,j, d(Z_{i,j},\hat{Z}_{i,j}) > \epsilon] \to 0$.  
\end{remark}

The following lemma shows that the lower bound for the classical problem without constant locality constraints also provides a lower bound for the problem here. 
\begin{lemma}
\label{lem:sideinformationouter}
We have
\begin{equation}
    R_{loc}(\epsilon) \geq \min_{\substack{U-S_1-S_2, S_1-S_2-V \\ \exists g: \mathbb{P}[d(Z,g(U,V))\leq \epsilon]=1}} I(U;S_1).
\end{equation}
\end{lemma}
\begin{IEEEproof}
The detailed proof can be found in Appendix~\ref{pflem:sideinformationouter}.
\end{IEEEproof}

The source $(S_1,S_2)$ is said to be {\it $\mathcal{S}_1$-regular} if there exists some $s_2 \in \mathcal{S}_2$ such that $p(s_1,s_2)>0, \forall s_1 \in \mathcal{S}_1$.
The following theorem shows that under a mild regularity condition, the achievable rate given by~\eqref{eq:R(e)} is tight.

\begin{theorem}
\label{thm:sideinformationexact}
If $(S_1,S_2)$ is $\mathcal{S}_1$-regular, then $R_{loc}(\epsilon) = \tilde{R}_{loc}(\epsilon)$.
\end{theorem}

\begin{IEEEproof}
The detailed proof can be found in Appendix~\ref{pfthm:sideinformationexact}.
\end{IEEEproof}

It is intuitive that the constant locality constraint impedes the decoding process, hence the optimal rate $ R_{loc}(\epsilon)$ is larger than that for the classical problem without constant locality.
More precisely, compared with the classical problem, the minimization constraints in~\eqref{eq:R(e)} do not change, while the objective function increases from $I(U;S_1|S_2)$ to $I(U;S_1)$.  \Cref{thm:sideinformationexact} implies that the decoder cannot efficiently use its knowledge of~$S_2^n$ to aid the decoding of~$U^n$. This effect is similar to the case where causal side information is available~\cite{Weissman2006}, and these two problems have the same optimal rate.

\Cref{thm:sideinformationexact} can be used to obtain a stronger converse for the distributed computing problem in some special cases, which induces relaxed conditions on source distributions. For example, we can easily obtain the following corollary for the approximate compression setting. It shows the optimality of separate coding for a larger class of sources than that in \Cref{cor:dac}.

\begin{corollary}
The conclusion of~\Cref{cor:dac} remains true if the source $(X_1,X_2)$ is both $\mathcal{X}_1$-regular and $\mathcal{X}_2$-regular.
\end{corollary}

%% file: 7.tex
\section{Graph Characterizations and Examples}
\label{sec:graph}
In this section, we develop graph charaterizations for the achievable rate $\tilde{R}_{loc}(\epsilon)$ in~\eqref{eq:R(e)} and the achievable rate region~$\tilde{\mathscr{R}}_{loc}(\epsilon)$ in~\eqref{eq:innerbound}. The characterizations can be used to simplify the computation of the rate (region). Examples are given in~\Cref{subsec:example}. 

\subsection{Computing with Side Information}


Consider the achievable rate $\tilde{R}_{loc}(\epsilon)$ given by~\eqref{eq:R(e)}. Note that for each random variable $U$, $p(u|s_1)$ is a transition probability. The intuition is to represent $s_1 \in \mathcal{S}$ and $u \in \mathcal{U}$ by vertices, and $p(u|s_1)$ as a flow on edges. The main concern is to find a proper graph such that the number of edges is small enough such that the reconstruction constraint $\mathbb{P}[d(f(x_1,x_2),\hat{z}) \leq \epsilon]=1$ is satisfied, and large enough such that the rate is minimized. To achieve this, we define the characteristic bipartite graph as follows.

\begin{definition}
\label{def:graphsideinformation}
First let $B(\hat{z},\epsilon) = \{z \in \mathcal{Z}| d(z,\hat{z})  \leq \epsilon\}$. Then let $\Gamma_{\epsilon}$ be the collection of all $u \subseteq{S}_1$ which satisfies the condition: for each $s_2 \in \mathcal{S}_2$, there exists some $\hat{z} \in \hat{\mathcal{Z}}$ such that, 
for any $s_1 \in u$ with $p(s_1,s_2)>0$, we have $f(s_1,s_2) \in B(\hat{z},\epsilon)$. 
Let $\mathcal{E}_{\epsilon} = \{(s_1,u)| s_1 \in u\}$. The {\it characteristic bipartite graph} for $\tilde{R}_{loc}(\epsilon)$ is defined as the  bipartite graph with two vertex sets $\mathcal{S}_1$, $\Gamma_{\epsilon}$ and an edge set $ \mathcal{E}_{\epsilon}$, denoted by $\mathcal{G}_{SI}[\mathcal{S}_1,\Gamma_{\epsilon}, \mathcal{E}_{\epsilon}]$.
\end{definition}

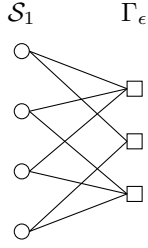
\begin{figure}[!t]
\centering
\begin{tikzpicture}
\node at (2,0.5) {$\mathcal{S}_1$};
\node at (3.5,0.5) {$\Gamma_{\epsilon}$};

\draw (2,0) circle(0.1);

\draw (2,-0.8) circle(0.1);

\draw (2,-1.6) circle(0.1);

\draw (2,-2.4) circle(0.1);

\draw (3.4,-0.6) rectangle (3.6,-0.4);

\draw (3.4,-1.3) rectangle (3.6,-1.1);

\draw (3.4,-2) rectangle (3.6,-1.8);

\draw  (2.1,0)--(3.4,-0.5);
\draw  (2.1,-0.8)--(3.4,-0.5);
\draw (2.1,-1.6)--(3.4,-0.5);

\draw (2.1,0)--(3.4,-1.2);
\draw (2.1,-2.4)--(3.4,-1.2);

\draw (2.1,-0.8)--(3.4,-1.9);
\draw (2.1,-1.6)--(3.4,-1.9);
\draw (2.1,-2.4)--(3.4,-1.9);
\end{tikzpicture}
\caption{An example of the characteristic bipartite graph.}
\label{fig:graph}
\vspace{-0.2cm}
\end{figure}

We see from above that each $s_1 \in \mathcal{S}_1$ must be covered by some $u \in \Gamma_{\epsilon}$ since $s_1 \in \{s_1\} \in \Gamma_{\epsilon}$. An example of the characteristic graph is shown in~\Cref{fig:graph}.
In light of~\Cref{def:graphsideinformation}, we have the following graph characterization for the achievable rate $\tilde{R}_{loc}(\epsilon)$, where ``$(S_1,U) \in \mathcal{E}_{\epsilon}$'' means $\mathbb{P}[(S_1,U) \in \mathcal{E}_{\epsilon}]=1$ and thus we omit the quantifier ``almost surely'' for simplicity.

\begin{theorem}
\label{thm:sideinformationgraph}
$\tilde{R}_{loc}(\epsilon) = \min_{p(u|s_1):(S_1,U) \in \mathcal{E}_{\epsilon}} I(U;S_1)$.
\end{theorem}

\begin{IEEEproof}
    The proof can be found in Appendix~\ref{pfthm:sideinformationgraph}.
\end{IEEEproof}

\begin{remark}
\label{rem:optimization}
Note that the optimization problem by~\Cref{thm:sideinformationgraph} is a special case of the unified source-channel optimization problem in~\cite{Yuan2023}, by letting $W$ be a constant and ignoring the loss constraint therein.
Therefore the algorithm in~\cite{Yuan2023} can be directly used for calculating the rate in~\Cref{thm:sideinformationgraph}.
Also note that the problem resembles the graph entropy problem in~\cite{Korner1973}. Though the graphs are constructed in different ways, these two optimization problems have the same structure.
\end{remark}

We can see from~\Cref{thm:sideinformationgraph} that the rate-tolerance function $\tilde{R}_{loc}(\epsilon)$ is a non-increasing, right-continuous step function. However, different from rate-distortion functions,  $\tilde{R}_{loc}(\epsilon)$ is generally not continuous in $\epsilon$. This can be seen from that the structure of the characteristic bipartite graph $\mathcal{G}_{SI}$ may change even if $\epsilon$ changes slightly, which can result in a significant change of the optimal value. 

Let 
$
    \Gamma_{\epsilon}^m = \{u \in \Gamma_{\epsilon}| u' \in \Gamma_{\epsilon}, u \subseteq u' \Rightarrow u = u' \}
$
be all maximal elements in $\Gamma_{\epsilon}$ under inclusion. Let $\mathcal{E}_{\epsilon}^m = \{(s_1,u)|s_1 \in u\}$. Then we construct a {\it maximal characteristic bipartite graph} $\mathcal{G}^m_{SI}[\mathcal{S}_1,\Gamma^m_{\epsilon}, \mathcal{E}^m_{\epsilon}]$.  We have the following corollary, which can be easily proved by combining~\Cref{thm:sideinformationgraph} and the data processing inequality.
\begin{corollary}
\label{cor:sideinformationgraph}
    $\tilde{R}_{loc}(\epsilon) = \min_{p(u|s_1):(S_1,U) \in \mathcal{E}^m_{\epsilon}} I(U;S_1)$.
\end{corollary}

Compared with~\Cref{thm:distributedgraph}, \Cref{cor:sideinformationgraph} reduces the size of the alphabet of $U$ and simplifies the computation of $\tilde{R}_{loc}(\epsilon)$. Examples are given in~\Cref{subsec:example}.

\subsection{The Distributed Computing Problem}

Now we consider the achievable rate region $\tilde{\mathscr{R}}_{loc}(\epsilon)$ in~\Cref{thm:distributedinner}. 
Different from the graph characterization for $\tilde{R}_{loc}(\epsilon)$, in this case there are generally more than one distributed characteristic bipartite graphs, each of which  consists of two graphs that should be constructed simultaneously.

\begin{definition}
\label{def:graphdistributed}
A {\it distributed characteristic bipartite hypergraph} $\mathcal{G}_{D}$ for $\tilde{\mathscr{R}}_{loc}(\epsilon)$ consists of two bipartite graphs $\mathcal{G}_1[\mathcal{X}_1,\Gamma_{1,\epsilon},\mathcal{E}_{1,\epsilon}]$ and $\mathcal{G}_2[\mathcal{X}_2,\Gamma_{2,\epsilon},\mathcal{E}_{2,\epsilon}]$ which satisfy,
\begin{enumerate}
    \item $\mathcal{X}_k$ is covered by $\Gamma_{k,\epsilon}$: for any $x_k \in \mathcal{X}_k$, there exists some $u_k \in \Gamma_{k,\epsilon}$ such that $x_k \in u_k$, $k = 1,2$;
    \item Elements in $\Gamma_{1,\epsilon} \times \Gamma_{2, \epsilon}$ are covered by $\epsilon$-balls: for any $u_1 \in \Gamma_{1,\epsilon}$ and $u_2 \in \Gamma_{2,\epsilon}$, there exists some $\hat{z} \in \hat{\mathcal{Z}}$ such that, for any $x_1 \in u_1$, $x_2 \in u_2$ with $p(x_1,x_2)>0$, we have $f(x_1,x_2) \in B(\hat{z},\epsilon)$;
    \item $\mathcal{E}_{k,\epsilon} = \{(x_k,u_k)|x_k \in u_k \}$, $k = 1,2$.
\end{enumerate}
\end{definition}

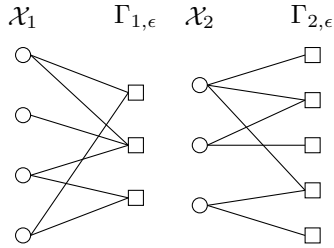
\begin{figure}[!t]
\centering
\begin{tikzpicture}
\node at (2,0.5) {$\mathcal{X}_1$};
\node at (3.5,0.5) {$\Gamma_{1,\epsilon}$};

\draw (2,0) circle(0.1);

\draw (2,-0.8) circle(0.1);

\draw (2,-1.6) circle(0.1);

\draw (2,-2.4) circle(0.1);

\draw (3.4,-0.6) rectangle (3.6,-0.4);

\draw (3.4,-1.3) rectangle (3.6,-1.1);

\draw (3.4,-2) rectangle (3.6,-1.8);

\draw  (2.1,0)--(3.4,-0.5);
\draw  (2.1,-2.4)--(3.4,-0.5);

\draw  (2.1,0)--(3.4,-1.2);
\draw  (2.1,-0.8)--(3.4,-1.2);
\draw (2.1,-1.6)--(3.4,-1.2);

\draw (2.1,-1.6)--(3.4,-1.9);
\draw (2.1,-2.4)--(3.4,-1.9);
\end{tikzpicture}
\begin{tikzpicture}
\node at (2,0.5) {$\mathcal{X}_2$};
\node at (3.5,0.5) {$\Gamma_{2,\epsilon}$};

\draw (2,-0.4) circle(0.1);

\draw (2,-1.2) circle(0.1);

\draw (2,-2) circle(0.1);

\draw (3.4,-0.1) rectangle (3.6,0.1);

\draw (3.4,-0.7) rectangle (3.6,-0.5);

\draw (3.4,-1.3) rectangle (3.6,-1.1);

\draw (3.4,-1.9) rectangle (3.6,-1.7);

\draw (3.4,-2.5) rectangle (3.6,-2.3);

\draw  (2.1,-0.4)--(3.4,-0);

\draw  (2.1,-0.4)--(3.4,-0.6);
\draw  (2.1,-1.2)--(3.4,-0.6);

\draw  (2.1,-1.2)--(3.4,-1.2);

\draw  (2.1,-0.4)--(3.4,-1.8);
\draw  (2.1,-2.0)--(3.4,-1.8);

\draw  (2.1,-2.0)--(3.4,-2.4);
\end{tikzpicture}
\caption{An example of the distributed characteristic bipartite graph.}
\label{fig:distributedgraph}
\vspace{-0.2cm}
\end{figure}

An example of distributed characteristic bipartite hypergraph~$\mathcal{G}_{D}$ is given in~\Cref{fig:distributedgraph}. 

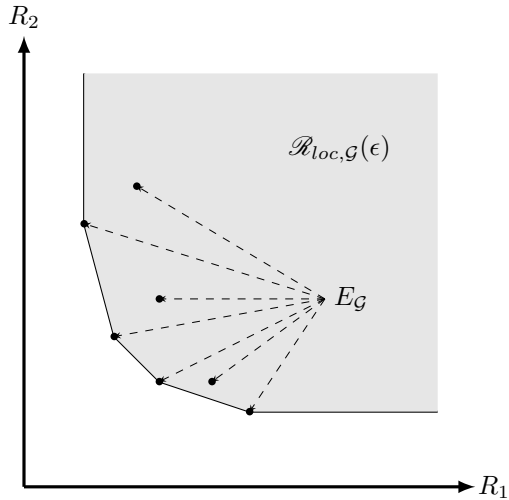
\begin{figure}[!t]
\centering
\begin{tikzpicture}
\draw[-latex,very thick] (0,0)--(6,0);
\node at (6.25,0) {$R_1$};
\draw[-latex,very thick] (0,0)--(0,6);
\node at (0,6.25) {$R_2$};

\draw[thick] (0.8,5.5)--(0.8,3.5)--(1.2,2)--(1.8,1.4)--(3,1)--(5.5,1);
\fill[gray!20] (0.8,5.5)--(0.8,3.5)--(1.2,2)--(1.8,1.4)--(3,1)--(5.5,1)--(5.5,5.5);

\fill (0.8,3.5) circle (0.05);
\fill (1.2,2) circle (0.05);
\fill (1.8,1.4) circle (0.05);
\fill (3,1) circle (0.05);

\fill (2.5,1.4) circle (0.05);
\fill (1.5,4) circle (0.05);
\fill (1.8,2.5) circle (0.05);
\draw[dashed,->,>=stealth,ultra thin] (4,2.5)--(2.5,1.4);
\draw[dashed,->,>=stealth,ultra thin] (4,2.5)--(1.5,4);
\draw[dashed,->,>=stealth,ultra thin] (4,2.5)--(1.8,2.5);

\node at (4.2,4.5) {$\mathscr{R}_{loc,\mathcal{G}}(\epsilon)$};

\draw[dashed,->,>=stealth,ultra thin] (4,2.5)--(0.8,3.5);
\draw[dashed,->,>=stealth,ultra thin] (4,2.5)--(1.2,2);
\draw[dashed,->,>=stealth,ultra thin] (4,2.5)--(1.8,1.4);
\draw[dashed,->,>=stealth,ultra thin] (4,2.5)--(3,1);
\node at (4,2.5) [right]{$E_{\mathcal{G}}$};
\end{tikzpicture}
\caption{The graph-based bound $\mathscr{R}_{loc,\mathcal{G}}(\epsilon)$.}
\label{fig:polygon}
\vspace{-0.2cm}
\end{figure}

Let $E_{\mathcal{G}}$ be the set of all points $(R_1,R_2)$ such that
\begin{equation}
\label{eq:graphinnerbound}
\begin{aligned}
    R_1  = \min_{p(u_1|x_1):(X_1, U_1) \in \mathcal{E}_{1,\epsilon}} I(X_1,U_1),
    \\
    R_2 = \min_{p(u_2|x_2):(X_2, U_2) \in \mathcal{E}_{2,\epsilon}} I(X_2;U_2),
\end{aligned}
\end{equation}
for some distributed characteristic bipartite graph $\mathcal{G}_{D}$ consisting of $\mathcal{G}_1[\mathcal{X}_1,\Gamma_{1,\epsilon},\mathcal{E}_{1,\epsilon}]$ and $\mathcal{G}_2[\mathcal{X}_2,\Gamma_{2,\epsilon},\mathcal{E}_{2,\epsilon}]$.
Let $\mathscr{R}_{loc,\mathcal{G}}(\epsilon) = \mathrm{conv} (E_{\mathcal{G}} \cup \{(+\infty,+\infty)\})$, 
which is an infinite polygon depicted in~\Cref{fig:polygon}.

We have the following characterization, which implies that $\tilde{\mathscr{R}}_{loc}(\epsilon)$ is an infinite polygon as well.
\begin{theorem}
\label{thm:distributedgraph}
$\tilde{\mathscr{R}}_{loc}(\epsilon) = \mathscr{R}_{loc,\mathcal{G}}(\epsilon)$. 
\end{theorem}

\begin{IEEEproof}
    The proof can be found in Appendix~\ref{pfthm:distributedgraph}.
\end{IEEEproof}

\begin{remark}
\label{rem:distributedoptimization}
Note that for each $\mathcal{G}_{D}$, the optimization problems for both rates in~\eqref{eq:graphinnerbound} can be independently solved.
Moreover, similar to the discussion in~\Cref{rem:optimization}, each of them is a special case of the unified source-channel optimization problem in~\cite{Yuan2023}. This makes the computation of the rate region much easier.  
\end{remark}

From~\Cref{thm:distributedgraph} and~\Cref{rem:distributedoptimization}, we see that it is easy to compute the rate region as long as all the graphs $\mathcal{G}_D$ are constructed. 
Moreover, similar to~\Cref{cor:sideinformationgraph}, we can simplify the computation by restricting the range of graphs been  searched.

Let $\mathcal{G}_{D}^j$, $j = 1,2$ be two distributed characteristic bipartite graphs, where $\mathcal{G}_{D}^j$ consists of $\mathcal{G}^j_1[\mathcal{X}_1,\Gamma^j_{1,\epsilon},\mathcal{E}^j_{1,\epsilon}]$ and $\mathcal{G}^j_2[\mathcal{X}_2,\Gamma^j_{2,\epsilon},\mathcal{E}^j_{2,\epsilon}]$. We define $\mathcal{G}_{D}^1 \preceq \mathcal{G}_{D}^2$ if
\begin{enumerate}
\item for any $k = 1,2$ and $u_k \in \Gamma^1_{k,\epsilon}$, there exists some $u_k' \in \Gamma^2_{k,\epsilon}$ such that $u_k \subseteq u_k'$;
\item $|\Gamma^1_{k,\epsilon}| \geq |\Gamma^2_{k,\epsilon}|$, $k = 1,2$.
\end{enumerate}

It is easy to verify that ``$\preceq$'' is a partial order. 
Note that if $\mathcal{G}_{D}^1 \preceq \mathcal{G}_{D}^2$ and $(R^j_1,R^j_2)$, $j = 1,2$ are corresponding rate pairs defined by~\eqref{eq:graphinnerbound}, then by the data processing inequality we have $R^1_1 \geq R^2_1$ and $R^1_2 \geq R^2_2$.
Let $E_{\mathcal{G}}^m$ be the set of rate pairs induced by distributed characteristic bipartite graphs that are maximal under the order $\preceq$. Then we have the following corollary of~\Cref{thm:distributedgraph}.

\begin{corollary}
\label{cor:distributedgraph}
$\tilde{\mathscr{R}}_{loc}(\epsilon) = \mathrm{conv} (E^m_{\mathcal{G}} \cup \{(\infty,\infty)\})$.
\end{corollary}

\Cref{cor:distributedgraph} is analogous to~\Cref{cor:sideinformationgraph}, but the definitions of graph maximality are different. In \Cref{cor:distributedgraph}, a maximal graph is defined over different graphs, while in \Cref{cor:sideinformationgraph} it is defined by maximal edges. Thus in \Cref{cor:distributedgraph}, we can not optimize over a single graph. 
By traversing all maximal distributed characteristic bipartite graphs, we can compute $E_{\mathcal{G}}^m$ and obtain $\tilde{\mathscr{R}}_{loc}(\epsilon)$.

Further note that for the problem with more than two encoders discussed in~\Cref{sec:more}, graph characterizations for the achievable region
$\tilde{\mathscr{R}}^{m}_{loc}(\epsilon)$ can be similarly developed.


\subsection{Examples}
\label{subsec:example}
In this subsection, we give some examples and calculate the achievable rate (region) using graph characterizations.

\begin{example}[Online card game]
\label{eg:card}
Alice and Bob each randomly select one out of three cards labeled 1, 2, and 3 without replacement.
Alice agrees to help Bob determine who selected the card with a larger label.
Consider the corresponding approximate computing problem with side information.
Denote the label of Alice's card  by $S_1$, and Bob's by $S_2$. 
Then  $(S_1,S_2) \sim p(s_1,s_2)$ with $p(i,j) = \frac{1}{6}(1-\delta_{i,j}),i,j=1,2,3$,
and $f(s_1,s_2) = \mathds{1}\{s_1>s_2\}$, where $\delta_{i,j}=1$ if $i=j$ and 0 otherwise. 
Let $d$ be the Hamming distortion on $\mathcal{Z} = \hat{\mathcal{Z}}  = \{ 0,1 \}$. 

We compute the achievable rate $\tilde{R}_{loc}(\epsilon)$ by the graph characterization in~\Cref{thm:sideinformationgraph}. 
\begin{enumerate}
\item For $\epsilon < 1$, we have $\Gamma^m_{\epsilon} = \{\{1,2\},\{2,3\}\}$ and then $\tilde{R}_{loc}(\epsilon) = \frac{2}{3}$.
\item For $\epsilon \geq 1$, it is easy to see that $\tilde{R}_{loc}(\epsilon) = R_{loc}(\epsilon) = 0$.
\end{enumerate}
\end{example}

\begin{example}[AND gate]
\label{eg:and}
Consider an distributed approximate computing problem. Let $(X_1,X_2) \sim \mathrm{DSBS}(p), 0 < p < 1$, and the function $f(X_1,X_2)$ is the AND gate of $X_1$ and $X_2$.
Let $d$ be the Hamming distortion on $\mathcal{Z} = \hat{\mathcal{Z}}  = \{ 0,1 \}$. 
By~\Cref{thm:distributedexact} and the graph characterization in~\Cref{thm:distributedgraph}, we compute the rate region~$\mathscr{R}_{loc}(\epsilon)$.
\begin{enumerate}
\item
For $\epsilon < 1$, there is only one maximal distributed characteristic bipartite graph with $\Gamma_{1,\epsilon} = \Gamma_{2,\epsilon} = \{ \{0\},\{1\}\}$, hence $I(X_k;U_k) = H(X_k), k = 1,2$. So the rate region is $\mathscr{R}_{loc}(\epsilon)=\{(R_1, R_2)|R_1 \geq 1, R_2 \geq 1\}$. 
\item For $\epsilon \geq 1$, there is only one maximal distributed characteristic graph with $\Gamma_{1,\epsilon} = \Gamma_{2,\epsilon} = \{ \{0,1\}\}$ and $\mathscr{R}_{loc}(\epsilon)=[0,\infty) \times [0,\infty)$, i.e., no communication is needed at all.
\end{enumerate}
\end{example}

\begin{figure}[!t]
\centering  
\begin{tikzpicture}[scale = 0.85]
\tikzstyle{every node}=[font=\small,scale=0.85]
\draw[-latex] (0,0)--(3,0);
\node at (3.25,0) {$R_1$};
\draw[-latex] (0,0)--(0,3);
\node at (0,3.25) {$R_2$};
\node at (-0.2,-0.2) {$0$};
\node at (1.585,-0.3) {$\log_2 3$};
\node at (0.667,-0.3) {$\frac{2}{3}$};
\node at (-0.5,1.585) {$\log_2 3$};
\node at (-0.3,0.667) {$\frac{2}{3}$};
\draw (1.585,0)--(1.585,0.1);
\draw (0.667,0)--(0.667,0.1);
\draw (0,1.585)--(0.1,1.585);
\draw (0,0.667)--(0.1,0.667);

\fill (1.585,1.585) circle (0.05);
\draw[dotted] (1.585,1.585)--(1.585,0);
\draw[dotted] (1.585,1.585)--(0,1.585);

\draw[] (1.585,2.75)--(1.585,1.585)--(2.75,1.585);
\fill[gray!20] (1.585,2.75)--(1.585,1.585)--(2.75,1.585)--(2.75,2.75);

\node at (1.6,-0.8) {$0 \leq \epsilon < \frac{1}{2}$};
\end{tikzpicture}
\begin{tikzpicture}[scale = 0.85]
\tikzstyle{every node}=[font=\small,scale=0.85]
\draw[-latex] (0,0)--(3,0);
\node at (3.25,0) {$R_1$};
\draw[-latex] (0,0)--(0,3);
\node at (0,3.25) {$R_2$};
\node at (-0.2,-0.2) {$0$};
\node at (1.585,-0.3) {$\log_2 3$};
\node at (0.667,-0.3) {$\frac{2}{3}$};
\node at (-0.5,1.585) {$\log_2 3$};
\node at (-0.3,0.667) {$\frac{2}{3}$};
\draw (1.585,0)--(1.585,0.1);
\draw (0.667,0)--(0.667,0.1);
\draw (0,1.585)--(0.1,1.585);
\draw (0,0.667)--(0.1,0.667);

\fill (0.667,1.585) circle (0.05);
\fill (1.585,0.667) circle (0.05);
\draw[dotted] (0.667,1.585)--(0,1.585);
\draw[dotted] (0.667,1.585)--(0.667,0);
\draw[dotted] (1.585,0.667)--(1.585,0);
\draw[dotted] (1.585,0.667)--(0,0.667);

\draw[] (0.667,2.75)--(0.667,1.585)--(1.585,0.667)--(2.75,0.667);
\fill[gray!20] (0.667,2.75)--(0.667,1.585)--(1.585,0.667)--(2.75,0.667)--(2.75,2.75);

\node at (1.6,-0.8) {$\frac{1}{2} \leq \epsilon < \frac{\sqrt{2}}{2}$};
\end{tikzpicture}

\begin{tikzpicture}[scale = 0.85]
\tikzstyle{every node}=[font=\small,scale=0.85]
\draw[-latex] (0,0)--(3,0);
\node at (3.25,0) {$R_1$};
\draw[-latex] (0,0)--(0,3);
\node at (0,3.25) {$R_2$};
\node at (-0.2,-0.2) {$0$};
\node at (1.585,-0.3) {$\log_2 3$};
\node at (0.667,-0.3) {$\frac{2}{3}$};
\node at (-0.5,1.585) {$\log_2 3$};
\node at (-0.3,0.667) {$\frac{2}{3}$};
\draw (1.585,0)--(1.585,0.1);
\draw (0.667,0)--(0.667,0.1);
\draw (0,1.585)--(0.1,1.585);
\draw (0,0.667)--(0.1,0.667);

\fill (0.667,0.667) circle (0.05);
\draw[dotted] (0.667,0.667)--(0.667,0);
\draw[dotted] (0.667,0.667)--(0,0.667);

\draw[] (0.667,2.75)--(0.667,0.667)--(2.75,0.667);
\fill[gray!20] (0.667,2.75)--(0.667,0.667)--(2.75,0.667)--(2.75,2.75);

\node at (1.6,-0.8) {$\frac{\sqrt{2}}{2} \leq \epsilon < 1$};
\end{tikzpicture}
\begin{tikzpicture}[scale = 0.85]
\tikzstyle{every node}=[font=\small,scale=0.85]
\draw[-latex] (0,0)--(3,0);
\node at (3.25,0) {$R_1$};
\draw[-latex] (0,0)--(0,3);
\node at (0,3.25) {$R_2$};
\node at (-0.2,-0.2) {$0$};
\node at (1.585,-0.3) {$\log_2 3$};
\node at (0.667,-0.3) {$\frac{2}{3}$};
\node at (-0.5,1.585) {$\log_2 3$};
\node at (-0.3,0.667) {$\frac{2}{3}$};
\draw (1.585,0)--(1.585,0.1);
\draw (0.667,0)--(0.667,0.1);
\draw (0,1.585)--(0.1,1.585);
\draw (0,0.667)--(0.1,0.667);

\fill (0,1.585) circle (0.05);
\fill (1.585,0) circle (0.05);
\fill (0.667,0.667) circle (0.05);
\draw[dotted] (0.667,0.667)--(0.667,0);
\draw[dotted] (0.667,0.667)--(0,0.667);

\draw[] (0,2.75)--(0,1.585)--(0.667,0.667)--(1.585,0)--(2.75,0);
\fill[gray!20] (0,2.75)--(0,1.585)--(0.667,0.667)--(1.585,0)--(2.75,0)--(2.75,2.75);

\node at (1.6,-0.8) {$1 \leq \epsilon < \frac{\sqrt{5}}{2}$};
\end{tikzpicture}

\begin{tikzpicture}[scale = 0.85]
\tikzstyle{every node}=[font=\small,scale=0.85]
\draw[-latex] (0,0)--(3,0);
\node at (3.25,0) {$R_1$};
\draw[-latex] (0,0)--(0,3);
\node at (0,3.25) {$R_2$};
\node at (-0.2,-0.2) {$0$};
\node at (1.585,-0.3) {$\log_2 3$};
\node at (0.667,-0.3) {$\frac{2}{3}$};
\node at (-0.5,1.585) {$\log_2 3$};
\node at (-0.3,0.667) {$\frac{2}{3}$};
\draw (1.585,0)--(1.585,0.1);
\draw (0.667,0)--(0.667,0.1);
\draw (0,1.585)--(0.1,1.585);
\draw (0,0.667)--(0.1,0.667);

\fill (0,0.667) circle (0.05);
\fill (0.667,0) circle (0.05);

\draw[] (0,2.75)--(0,0.667)--(0.667,0)--(2.75,0);
\fill[gray!20] (0,2.75)--(0,0.667)--(0.667,0)--(2.75,0)--(2.75,2.75);

\node at (1.6,-0.8) {$\frac{\sqrt{5}}{2} \leq \epsilon < \sqrt{2}$};
\end{tikzpicture}
\begin{tikzpicture}[scale = 0.85]
\tikzstyle{every node}=[font=\small,scale=0.85]
\draw[-latex] (0,0)--(3,0);
\node at (3.25,0) {$R_1$};
\draw[-latex] (0,0)--(0,3);
\node at (0,3.25) {$R_2$};
\node at (-0.2,-0.2) {$0$};
\node at (1.585,-0.3) {$\log_2 3$};
\node at (0.667,-0.3) {$\frac{2}{3}$};
\node at (-0.5,1.585) {$\log_2 3$};
\node at (-0.3,0.667) {$\frac{2}{3}$};
\draw (1.585,0)--(1.585,0.1);
\draw (0.667,0)--(0.667,0.1);
\draw (0,1.585)--(0.1,1.585);
\draw (0,0.667)--(0.1,0.667);

\fill (0,0) circle (0.05);

\draw[] (0,2.75)--(0,0)--(2.75,0);
\fill[gray!20] (0,2.75)--(0,0)--(2.75,0)--(2.75,2.75);

\node at (1.6,-0.8) {$\epsilon \geq \sqrt{2}$};
\end{tikzpicture}
\caption{The rate region $\mathscr{R}_{loc}(\epsilon)$ for \Cref{eg:Euc}.}
\label{fig:eg}
\vspace{-0.2cm}
\end{figure}
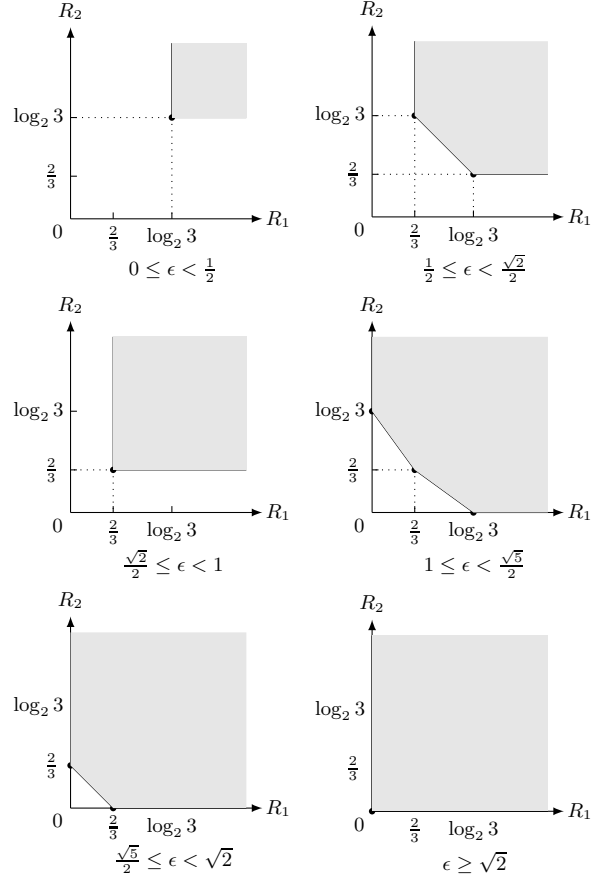

\begin{example}[Distributed approximate compression under Euclidean distance]
\label{eg:Euc}
Let $\mathcal{X}_1 = \mathcal{X}_2 = \{1,2,3\}$, $p(x_1,x_2) = \frac{1}{9}, \forall x_1,x_2$, and $f(X_1,X_2)=(X,Y)$.
Let $d$ be the Euclidean distance, and $\hat{\mathcal{Z}}  = \mathbb{R}^2$. We use~\Cref{thm:distributedexact} and the graph charaterization by~\Cref{thm:distributedgraph} to compute the rate region $\mathscr{R}_{loc}(\epsilon)$.
Note that by~\Cref{cor:distributedgraph}, we only need to compute all points in $E_{\mathcal{G}}^m$, which are given as follows. The corresponding rate regions are plotted in~\Cref{fig:eg}.
\begin{enumerate} \item For $0 \leq \epsilon < \frac{1}{2}$, we have $|E^m_{\mathcal{G}}| = 1$.
$\Gamma_{1,\epsilon}^1 = \Gamma_{2,\epsilon}^1 = \{ \{1\},\{2\},\{3\}\}$ induces a rate pair $(\log_2 3,\log_2 3)$.
\item For $\frac{1}{2} \leq \epsilon < \frac{\sqrt{2}}{2}$, we have $|E^m_{\mathcal{G}}| = 2$.
$\Gamma_{1,\epsilon}^1 = \{\{1,2\},\{2,3\}\}, \Gamma_{2,\epsilon}^1 = \{ \{1\},\{2\},\{3\}\}$ induces $(\frac{2}{3},\log_2 3)$. $\Gamma_{1,\epsilon}^2 = \{ \{1\},\{2\},\{3\}\}, \Gamma_{2,\epsilon}^2 = \{\{1,2\},\{2,3\}\}$ induces $(\log_2 3,\frac{2}{3})$. 
\item For $\frac{\sqrt{2}}{2} \leq \epsilon < 1$, we have $|E^m_{\mathcal{G}}| = 1$. $\Gamma_{1,\epsilon}^1 =\Gamma_{2,\epsilon}^1 =  \{\{1,2\},\{2,3\}\}$ induces $(\frac{2}{3},\frac{2}{3})$.
\item For $1 \leq \epsilon < \frac{\sqrt{5}}{2}$, we have $|E^m_{\mathcal{G}}| = 3$. $\Gamma_{1,\epsilon}^1 =\Gamma_{2,\epsilon}^1 =  \{\{1,2\},\{2,3\}\}$ induces $(\frac{2}{3},\frac{2}{3})$.
$\Gamma_{1,\epsilon}^2 = \{ \{1\},\{2\},\{3\}\}, \Gamma_{2,\epsilon}^2 = \{\{1,2,3\}\}$ induces $(\log_2 3,0)$.
$\Gamma_{1,\epsilon}^3 = \{\{1,2,3\}\}, \Gamma_{2,\epsilon}^3 = \{ \{1\},\{2\},\{3\}\}$ induces $(0,\log_2 3)$.
\item For $\frac{\sqrt{5}}{2} \leq \epsilon < \sqrt{2}$, we have $|E^m_{\mathcal{G}}| = 2$.
$\Gamma_{1,\epsilon}^1 = \{\{1,2\},\{2,3\}\}, \Gamma_{2,\epsilon}^1 = \{ \{1,2,3\}\}$ induces $(\frac{2}{3},0)$. $\Gamma_{1,\epsilon}^2 = \{ \{1,2,3\}\}, \Gamma_{2,\epsilon}^2 = \{\{1,2\},\{2,3\}\}$ induces $(0,\frac{2}{3})$. 
\item For $\epsilon \geq \sqrt{2}$, we have $|E^m_{\mathcal{G}}| = 1$.
$\Gamma_{1,\epsilon}^1 = \Gamma_{2,\epsilon}^1 = \{ \{1,2,3\}\}$ induces $(0,0)$.
\end{enumerate}
\end{example}

%% file: 8.tex
\section{Conclusion}
\label{sec:conclusion}
In this work, we aimed at establishing the rate region for the distributed coding for computing problem with constant decoding locality. 
We designed an efficient layered coding scheme by taking advantages of both classic typicality coding techniques and an expander graph code.
Then we proved the optimality of the induced rate region under some regularity conditions on sources.
Compared with the classical problem without constant locality constraints, the rate region for the problem here is strictly smaller in most cases. 
Our results characterized the tradeoff between communication costs and coding complexity for the distributed computing problem.

It is natural that the tradeoff can be similarly characterized for variants of the problem.
Both the achievable and the converse parts were generalized to cases with more than two encoders. 
The coding for computing problem with side information was analogously solved.

Furthermore, we developed graph characterizations for the above 
rate regions. By constructing distributed characteristic bipartite graphs, the computation of the rate region becomes much easier both analytically and numerically.

Directions for future work include determining the exact rate region for general sources, and analyzing the tradeoff in other settings.
And we hope related results can provide insights for the design and analysis of practical systems.

%% file: A1.tex
\appendices

\section{Proof of Lemma~\ref{lem:partial}}
\label{pflem:partial}

For any $x_2 \in \mathcal{X}_2$, choose some $u_{2,x_2}$ from $\{u_2 \in \mathcal{U}_2| p(x_2,u_2)>0\}$ arbitrarily. For any $u_1 \in \mathcal{U}_1$ and $x_2 \in \mathcal{X}_2$, let $g_1(u_1,x_2) = g(u_1,u_{2,x_2})$.
Then for any $(x_1,x_2,u_1)$ with $p(x_1,x_2,u_1)>0$, by the Markov chain $U_1-X_1-X_2-U_2$ we have 
$p(x_1,x_2,u_1,u_{2,x_2}) = p(x_1,x_2,u_1)p(u_{2,x_2}|x_2)>0$.
Since $\mathbb{P}[d(f(X_1,X_2),g(U_1,U_2)) \leq \epsilon] = 1$, we have  $d(f(x_1,x_2),g_1(u_1,x_2))=d(f(x_1,x_2),g(u_1,u_{2,x_2})) \leq \epsilon$.
Thus we have $\mathbb{P}[d(f(X_1,X_2),g_1(U_1,X_2)) \leq \epsilon] = 1$.
The existence of $g_2$ can be shown similarly.

\section{Proof of Lemma~\ref{lem:distributedouter}}
\label{pflem:distributedouter}

Similar to~\Cref{lem:averagerate}, for any $(n,2^{nR_1},2^{nR_2},t)$ distributed code, it is easy to verify 
\begin{equation}
\label{eq:rates}
\begin{aligned}
R_1 \geq & \frac{1}{n}\sum_{i =1}^n I(X_{1,i};M_{1,I_{1,i}}),
\\
R_2 \geq & \frac{1}{n}\sum_{i =1}^n I(X_{2,i};M_{2,I_{2,i}}),
\\
R_1+R_2 \geq & \frac{1}{n}\sum_{i =1}^n I(X_{1,i},X_{2,i};M_{1,I_{1,i}},M_{2,I_{2,i}}).
\end{aligned}
\end{equation}
Let $U_{1,i} = M_{1,I_{1,i}}$, $U_{2,i} = M_{2,I_{2,i}}$. We introduce a time-sharing variable $Q \sim \mathrm{Unif}[1:n]$ independent of $(X_1^n,X_2^n)$, and let $X_1^{(n)} = X_{1,Q}$, $X_2^{(n)} = X_{2,Q}$, $U_1^{(n)} = (U_{1,Q},Q)$, $U_2^{(n)} = (U_{2,Q},Q)$ and $\hat{Z}^{(n)} = \hat{Z}_Q$. Then we have $\hat{Z}^{(n)} $ is a function of $(U_1^{(n)},U_2^{(n)})$,  and Markov chains $U_{1}^{(n)}-S_{1}^{(n)}-S_{2}^{(n)}$ and $S_{1}^{(n)}-S_{2}^{(n)}-U_{2}^{(n)}$ hold.
By~\eqref{eq:rates}, we have 
\begin{align*}
R_1 \geq & I(X_1^{(n)};U_1^{(n)}),
\\
R_2 \geq & I(X_2^{(n)};U_2^{(n)}),
\\
R_1+R_2 \geq &(X_1^{(n)},X_2^{(n)};U_1^{(n)},U_2^{(n)}).
\end{align*}
Let $n \to \infty$, then we complete the proof.

\section{Proof of Lemma~\ref{lem:averagerate}}
\label{pflem:averagerate}
Since the source $X_k$ is memoryless, $k = 1,2$, we have 
\begin{align*}
    nR_k =& H(M^{nR_k}_k) \geq I(X_k^n;M^{nR_k}_k) 
    \\
    =& \sum_{i = 1}^n I(X_{k,i};M^{nR_k}_k|X_{k}^{i-1})
    \\
    =& \sum_{i = 1}^n I(X_{k,i};M^{nR_k}_k,X_k^{i-1})
    \\
    \geq &\sum_{i = 1}^n I(X_{k,i};M_{k,I_{k,i}}),
\end{align*}
completing the proof.

\section{Proof of Corollary~\ref{cor:dac}}
\label{pfcor:dac}
By~\Cref{thm:distributedexact}, we have $\mathscr{R}_{loc}(\epsilon) = \tilde{\mathscr{R}}_{loc}(\epsilon)$.
Note that both $\tilde{\mathscr{R}}_{loc}(\epsilon)$ (c.f.~\eqref{eq:innerbound}) and the bound defined by~\eqref{eq:dacbound} are convex, hence it suffices to show their extreme point sets are the same. 
Define two sets of random variables to be 
\[
\mathscr{U}_1 = \left\{(U_1,U_2)\left|\begin{array}{c}
        U_1-X_1-X_2-U_2,\  \exists g_1,g_2,\\
    \mathbb{P}[d(X_1,g_1(U_1,U_2)) \leq \epsilon] = 1, \\
    \mathbb{P}[d(X_2,g_2(U_1,U_2)) \leq \epsilon] = 1, 
\end{array}\right.\right\}
\]
and 
\[
\mathscr{U}_2 = \left\{(U_1,U_2)\left|\begin{array}{c}
        U_1-X_1-X_2-U_2,\  \exists \tilde{g}_1,\tilde{g}_2,\\
    \mathbb{P}[d(X_1,\tilde{g}_1(U_1)) \leq \epsilon] = 1, \\
    \mathbb{P}[d(X_2,\tilde{g}_2(U_2)) \leq \epsilon] = 1. 
\end{array}\right.\right\}
\]
Then we only need to show that $\mathscr{U}_1 = \mathscr{U}_2$. 
Since one of the direction $\mathscr{U}_2 \subseteq \mathscr{U}_1$ is direct, it remains to show that $\mathscr{U}_1 \subseteq \mathscr{U}_2$.

Let $(U_1,U_2) \in \mathscr{U}_1$, and without loss of generality, we assume that $p(u_1)>0, \forall u_1 \in \mathcal{U}_1$ and $p(u_2)>0, \forall u_2 \in \mathcal{U}_2$. For any $u_1 \in \mathcal{U}_1$, choose $u_2 \in \mathcal{U}_2$ arbitrarily and define $\tilde{g}_1(u_1) = g_1(u_1,u_2)$.

Let $(x_1,u_1)$ satisfy $d(x_1,\tilde{g}_1(u_1)) > \epsilon$. Then there exists some $u_2 \in \mathcal{U}_2$ such that $d(x_1,g_1(u_1,u_2)) > \epsilon$.
Since $\mathbb{P}[d(X_1,g_1(U_1,U_2)) \leq \epsilon] = 1$, we have $p(u_1,x_1,u_2) = 0$. By the assumption that $p(u_2) >0$, there exists some $x_2 \in \mathcal{X}_2$ with $p(x_2,u_2)>0$. Since $(X_1,X_2)$ has a full support, we have $p(x_1|x_2)>0$. Then by the Markov chain $X_1-X_2-U_2$ we have $p(x_1,u_2) \geq p(x_1,x_2,u_2) = p(x_1|x_2)p(x_2,u_2)>0$.
Because of the Markov chain $U_1-X_1-U_2$, we further have $p(u_1,x_1,u_2) =p(u_1|x_1) p(x_1,u_2)$. This implies $p(u_1|x_1) = 0$ and then $p(u_1,x_1) = 0$. Since $(x_1,u_1)$ is arbitrary, we have proved that $\mathbb{P}[d(X_1,\tilde{g}_1(U_1)) \leq \epsilon] = 1$.

Similarly, $\tilde{g}_2$ can be constructed and we can prove that $\mathbb{P}[d(X_2,\tilde{g}_2(U_2)) \leq \epsilon] = 1$. Hence we have $(U_1,U_2) \in \mathscr{U}_2$. Then we can conclude that $\mathscr{U}_1 = \mathscr{U}_2$, which completes the proof.

%% file: A2.tex
\section{Proof of Theorem~\ref{thm:sideinformationinner}}
\label{pfthm:sideinformationinner}
Let $(U,g)$ achieve the minimum in the right hand of~\eqref{eq:R(e)}. We show the achievability by defining a series of $(nb,2^{nbR},t)$ code, where $b$, $R$ and $t$ are chosen in the following detailed proof.

\subsection{Layered Coding Scheme}
\subsubsection{Codebook Generation}
Let $R_1 = I(U;S_1) + \delta'$, then by the proof of the rate-distortion theorem (c.f.~\cite{EIGamal2011}), there exists some $b \in \mathbb{N}$ such that there exists a code $\mathcal{C}_1$  which consists of
an encoding function 
\[
h_{1,e}: \mathcal{S}_1^b \to \{0,1\}^{bR_1},
\]
and a decoding function
\[
h_{1,d}:\{0,1\}^{bR_1} \to \mathcal{U}^b,
\]
satisfying $\mathbb{P}\left[(h_{1,d}(h_{1,e}(S_{1}^b)),S_{1}^b) \notin \mathcal{T}^{b,\delta''}_{(U,S_1)}\right] = \delta$ for some $\delta''>0$.

Let $R_2 = \log |\mathcal{S}_1|+1$,  then there is a simple code $\mathcal{C}_2$ consisting of an encoding function 
\[
h_{2,e}: \mathcal{S}_1^b \to \{0,1\}^{bR_2},
\]
and a decoding function
\[
h_{2,d}:\{0,1\}^{bR_2} \to \mathcal{S}_1^b,
\]
such that for any $s_1^b \in \mathcal{S}_1^b$, $h_{2,e}(s_1^b) \neq 0^{b R_2}$ and $h_{2,d}(h_{2,e}(s_1^b)) = s_1^b$.

For any $n \in \mathbb{N}$, there exists a $(nbR_2, 2\delta, R_3 = O(\delta \log \frac{1}{\delta}),t = O(\log \frac{1}{\delta}))$ expander code $\mathcal{C}_3$ satisfying the properties in~\Cref{lem:expander}.
Let $R = R_1+R_2 R_3$. Assume the decoding function $h_{1,d}$ is known to the encoder.

\subsubsection{Encoding}
Given $S_{1}^{nb} = \big(S_{1,i}^b\big)_{i = 1}^n$, first compute $ M^{nbR_1}_1 = \big(M^{bR_1}_{1,i}\big)_{i = 1}^n= \big(h_{1,e}(S_{1,i}^b)\big)_{i = 1}^n$, $U^{nb} = \big(U_{i}^b\big)_{i = 1}^n = \big(h_{1,d}(M^{bR_1}_{1,i})\big)_{i = 1}^n$, and $\mathcal{J} = \{i:(U_{i}^b,S_{1,i}^b) \notin \mathcal{T}^{b,\delta''}_{(U,S_1)} \}$.
Then let $M^{nbR_2}_2 = \big(M^{bR_2}_{2,i}\big)_{i = 1}^n$, where 
\begin{equation}
    M^{bR_2}_{2,i} = \left\{
    \begin{array}{cc}
         h_{2,e}(S_{1,i}^b)& i \in \mathcal{J}.  
         \\
         0^{bR_2}& i \notin \mathcal{J}.  
    \end{array}\right.
\end{equation}
After that, encode $M^{nbR_2}_2$ by the coding scheme of $\mathcal{C}_3$ into $M^{nbR_2R_3}_3$. 
Finally, let the encoded message be $M^{nbR} = (M^{nbR_1}_1,M^{nbR_2R_3}_3)$, and send it to the decoder.

\subsubsection{Decoding}
Given $i \in \{1,...,n\}$ and $j \in \{1,...,b\}$, the decoder constructs $\hat{Z}_{i,j}$ as follows.
First decode $M^{bR_2}_{2,i}$ by probing some bits in $M^{nbR_2R_3}_3$.
\begin{enumerate}[(i)]
\item
If $M^{bR_2}_{2,i} = 0^{bR_2}$, then 
compute $U_{i}^b = h_{1,d}(M^{bR_1}_{1,i})$ and let $\hat{Z}_{i,j} = g(U_{i,j},S_{2,i,j})$;
\item
If $M^{bR_2}_{2,i} \neq 0^{bR_2}$, then compute $S_{1,i}^b = h_{2,d}(M^{bR_2}_{2,i})$ and let $\hat{Z}_{i,j}=f(S_{1,i,j},S_{2,i,j})$.
\end{enumerate}

\subsection{Analysis}
\subsubsection{Code Length Analysis}
The rate of $M^{nbR}$ is 
\[
R = R_1+R_2R_3= I(U;S_1)+ \delta'+O\left(\delta \log \frac{1}{\delta}\right),
\]
which can be arbitrarily close to $I(U;S_1)$ by letting $\delta'$ and $\delta$ be small enough.

\subsubsection{Number of Probed Bits}
The number of total probed bits for decoding $\hat{Z}_{i,j}$ is bounded by
\[
 bR_1 + bR_2t = b\left[I(U;S_1)+ \delta'+O\left(\log \frac{1}{\delta}\right)\right],
\]
which is a constant independent of $n$.

\subsubsection{Error Analysis}
Finally, we prove~\eqref{eq:achivable} holds as follows.
Let 
\begin{equation}
    E_i = \mathds{1}\{(U_{i}^b,S_{1,i}^b) \notin \mathcal{T}^{b,\delta''}_{(U,S_1)} \}, i \in \{1,...,n\}.
\end{equation}
Then $E_i \sim \mathrm{Bern}(\delta)$ are i.i.d. random variables. We have $\mathcal{J} = \{i: E_i = 1\}$ and by~\Cref{lem:Chernoff}, 
\begin{equation}
\label{eq:E_1}
    \mathbb{P}\left[|\mathcal{J}|> 2n\delta\right] \leq e^{-\frac{n\delta}{3}}.
\end{equation}
Let $\mathcal{E}_1 = \{|\mathcal{J}| > 2n\delta\}$, then for the event $\mathcal{E}_1^c$, we have $|\mathcal{J}| \leq 2n \delta$ and by~\Cref{lem:expander} the docoding result of $M^{bR_2}_{2,i}$ is always correct for $i \in\{1,...,n\}$. In this case, we further show the final reconstruction~$\hat{Z}_{i,j}$ satisfies $\mathbb{P}[d(Z_{i,j},\hat{Z}_{i,j}) \leq \epsilon|\mathcal{E}_1^c] = 1$ for any $i \in\{1,...,n\}$, $j \in \{1,...,b\}$, as follows.

\begin{enumerate}[(a)]
\item
For the event $\mathcal{E}_1^c  \cap \{i \notin \mathcal{J}\} $, we have $(U_{i}^b,S_{1,i}^b) \in \mathcal{T}^{b,\delta''}_{(U,S_1)}$. Then by~\eqref{eq:robusttypical} we have $p(u,s_1) > 0$ for $(U_{i,j},S_{1,i,j}) = (u,s_1)$ and $S_{2,i,j} = s_2$.
Hence we have $p(u,s_1,s_2) = p(u,s_1)p(s_2|s_1)>0$. Since $(U,g)$ satisfies that $\mathbb{P}[d(f(S_1,S_2),g(U,S_2))\leq \epsilon]=1$, we have $d(f(s_1,s_2),g(u,s_2))) \leq \epsilon$. By the case (i) in the decoding process, we have $\hat{Z}_{i,j} = g(U_{i,j},S_{2,i,j})$, thus $d(Z_{i,j}, \hat{Z}_{i,j})\leq \epsilon$.
\item For the event $\mathcal{E}_1^c \cap \{i \in \mathcal{J}\}$, by the case (ii) in the decoding process, we have $d(Z_{i,j}, \hat{Z}_{i,j})=0$. 
\end{enumerate}
Thus by~\eqref{eq:E_1},
\[
\mathbb{P}[d(Z_{i,j}, \hat{Z}_{i,j})> \epsilon] \leq \mathbb{P}[\mathcal{E}_1] \to 0, n \to \infty,
\]
which completes the proof.


%% file: A3.tex
\section{Proof of Lemma~\ref{lem:sideinformationouter}}
\label{pflem:sideinformationouter}
Similar to~\Cref{lem:averagerate}, For any $(n,2^{nR},t)$ code, we have 
\begin{equation}
\label{eq:29}
    R \geq \frac{1}{n}\sum_{i =1}^n I(S_{1,i};M_{I_i}).
\end{equation}
Let $U_{i} = M_{I_{i}}$, $V_{i} = S_2^n$. 
We introduce a time-sharing variable $Q \sim \mathrm{Unif}[1:n]$ independent of $(S_1^n,S_2^n)$, and let $S_1^{(n)} = S_{1,Q}$, $S_2^{(n)} = S_{2,Q}$, $U^{(n)} = (U_{Q},Q)$, $V^{(n)} = V_{Q}$ and $\hat{Z}^{(n)} = \hat{Z}_Q$. Then $\hat{Z}^{(n)}$ is a function of $(U^{(n)},V^{(n)})$,  and Markov chains $U^{(n)}-S_{1}^{(n)}-S_{2}^{(n)}$ and $S_{1}^{(n)}-S_{2}^{(n)}-V^{(n)}$ hold.
Then by~\eqref{eq:29}, we have 
\begin{align*}
R \geq & \frac{1}{n}\sum_{i =1}^n I(S_{1,i};U_{i})
\\
= &  I(S_{1,Q};U_{Q}|Q)
\\
= &  I(S_{1,Q};U_{Q},Q)
\\
= &I(S_1^{(n)};U^{(n)}).
\end{align*}
Let $n \to \infty$, then we complete the proof.

%% file: A4.tex
\section{Proof of Theorem~\ref{thm:sideinformationexact}}
\label{pfthm:sideinformationexact}

We first present the following lemma, which is useful in the proof of~\Cref{thm:sideinformationexact}.

\begin{lemma}
\label{lem:relation}
    If 
    \begin{equation}
    \label{eq:ratebound}
        R <  \min_{\substack{U-S_1-S_2, \\ \exists g: \mathbb{P}[d(Z,g(U,S_2))\leq \epsilon]=1}} I(U;S_1),
    \end{equation}
    then there exists some $i \in \{1,...,n\}$, $m \in \{0,1\}^{|I_i|}$ and $s_2^* \in \mathcal{S}_2$, such that for any $\hat{z} \in \mathcal{\hat{Z}}$,
    \begin{equation}
    \label{eq:relation}
    \mathbb{P}[d(Z_i, \hat{z}) > \epsilon, M_{I_i} = m, S_{2,i} = s_{2}^*] \geq \frac{1}{|\mathcal{S}_2|} 2^{-t}\delta.
\end{equation}
\end{lemma}

\begin{IEEEproof}[Proof of~\Cref{lem:relation}]
For any $(n,2^{nR},t)$ code, we have a Markov chain $M_{I_i}-S_{1,i}-S_{2,i}$.
By the proof of ~\Cref{lem:sideinformationouter}, 
there exists some $i \in \{1,...,n\}$ such that $I(S_{1,i};M_{I_i}) \leq R$. So by the condition in~\eqref{eq:ratebound},
we have 
\[
I(M_{I_i};S_{1,i}) < \min_{\substack{U-S_1-S_2, \\ \exists g: \mathbb{P}[d(Z,g(U,S_2))\leq \epsilon]=1}} I(U;S_1),
\]
which implies that $M_{I_i}$ does not satisfy the constraint for $U$.
Since the Markov chain holds, then there must exist some $\delta>0$ 
such that for any $g: \{0,1\}^{|I_i|} \times \mathcal{S}_2 \to \mathcal{\hat{Z}}$,
\[
\mathbb{P}[d(Z_i, g(M_{I_i},S_{2,i})) > \epsilon] \geq \delta.
\]
Hence there exist some $m \in \{0,1\}^{|I_i|}$ and $s_{2}^* \in \mathcal{S}_2$, such that 
\begin{equation*}
    \mathbb{P}[d(Z_i, g(m,s_{2}^*)) > \epsilon, M_{I_i} = m,S_{2,i} = s_{2}^*] \geq \frac{1}{2^{|I_i|}\cdot|\mathcal{S}_2|} \delta \geq \frac{1}{|\mathcal{S}_2|} 2^{-t}\delta.
\end{equation*}
Since the choice of $g(m,s_2^*)$ is arbitrary, we have for any $\hat{z} \in \mathcal{\hat{Z}}$, 
\begin{equation*}
    \mathbb{P}[d(Z_i, \hat{z}) > \epsilon, M_{I_i} = m, S_{2,i} = s_{2}^*] \geq \frac{1}{|\mathcal{S}_2|} 2^{-t}\delta.
\end{equation*}
\end{IEEEproof}

Next we prove~\Cref{thm:sideinformationexact}. By~\Cref{thm:sideinformationinner}, it remains to show $R \geq  \tilde{R}_{loc}(\epsilon)$.
We show by contradiction. For any $(n,2^{nR},t)$ code, suppose $
        R <  \tilde{R}_{loc}(\epsilon)$.
By~\Cref{lem:relation}, there exists some $i \in \{1,...,n\}$, $m \in \{0,1\}^{|I_i|}$ and $s_2^* \in \mathcal{S}_2$, such that for any $\hat{z} \in \mathcal{\hat{Z}}$,
\[
    \mathbb{P}[d(Z_i, \hat{z}) > \epsilon, M_{I_i} = m, S_{2,i} = s_{2}^*] \geq \frac{1}{|\mathcal{S}_2|} 2^{-t}\delta.
\]
Note that 
\[
\mathbb{P}[d(Z_i, \hat{z}) > \epsilon, M_{I_i} = m, S_{2,i} = s_{2}^*] = \sum_{s_1: d(f(s_1,s_2^*),\hat{z})> \epsilon} \mathbb{P}[ M_{I_i} = m,S_{1,i} =s_{1},S_{2,i} = s_{2}^*].
\]
Then there exists some $s_{1,\hat{z}} \in \mathcal{S}_1$ such that 
\begin{equation}
\label{eq:dddd}
d(f(s_{1,\hat{z}},s_{2}^*),\hat{z}) > \epsilon
\end{equation}
and 
\begin{equation}
\label{eq:lowerprob}
\mathbb{P}[ M_{I_i} = m,S_{1,i} =s_{1,\hat{z}},S_{2,i} = s_{2}^*] \geq \frac{1}{|\mathcal{S}_1||\mathcal{S}_2|}2^{-t}\delta.
\end{equation}

Define the coupling $((S^n_{1,\hat{z}})_{\hat{z} \in \mathcal{\hat{Z}}},S_2^n)$ by letting $p((s^n_{1,\hat{z}})_{\hat{z} \in \mathcal{\hat{Z}}},s_2^n) = p_{S_2^n}(s_2^n) \prod_{\hat{z}}p_{S_1^n|S_2^n}(s^n_{1,\hat{z}} |s_2^n)$, then $(Z^n_{\hat{z}})_{\hat{z} \in \mathcal{\hat{Z}}}$ and $(M^{nR}_{\hat{z}})_{\hat{z} \in \mathcal{\hat{Z}}}$ are induced naturally. 
Then we have
\begin{align}
         &\mathbb{P}[d(Z_i,\hat{Z}_i)> \epsilon]
         \geq \mathbb{P}[d(Z_i,\hat{Z}_i)> \epsilon, M_{I_i} = m] 
         \nonumber \\
         = &\sum_{s_2^n} \mathbb{P}[S_2^n = s_2^n] \mathbb{P}[d(Z_i,\hat{Z}_i)> \epsilon, M_{I_i} = m | S_2^n = s_2^n] 
         \nonumber \\
         \geq & \sum_{s_2^n}\mathbb{P}[S_2^n = s_2^n] \min_{\hat{z} \in \mathcal{\hat{Z}}} \mathbb{P}[d(Z_i, \hat{z}) >\epsilon, M_{I_i} = m|S_2^n = s_2^n] 
         \nonumber \\
         \geq& \sum_{s_2^n}\mathbb{P}[S_2^n = s_2^n] \prod_{\hat{z} \in \mathcal{\hat{Z}}} \mathbb{P}[d(Z_i, \hat{z}) >\epsilon, M_{I_i} = m | S_2^n = s_2^n] 
         \nonumber \\
         =  & \sum_{s_2^n}\mathbb{P}[S_2^n = s_2^n] \mathbb{P}\left[\cap_{\hat{z} \in \mathcal{\hat{Z}}} \{d(Z_{\hat{z},i}, \hat{z}) >\epsilon, M_{\hat{z},I_i} = m  | S_2^n = s_2^n\}\right]
         \nonumber \\
         =  & \mathbb{P}\left[\cap_{\hat{z} \in \mathcal{\hat{Z}}} \{d(Z_{\hat{z},i}, \hat{z}) >\epsilon, M_{\hat{z},I_i} = m \}\right]
         \nonumber \\
         \geq & \mathbb{P}\left[\cap_{\hat{z} \in \mathcal{\hat{Z}}} \{d(Z_{\hat{z},i}, \hat{z}) >\epsilon, M_{\hat{z},I_i} = m \} \cap \{S_{2,i} = s_2^*\}\right]
         \label{eq:last0} \\
         \geq & \mathbb{P}\left[\cap_{\hat{z} \in \mathcal{\hat{Z}}} \{S_{1,\hat{z},i} =s_{1,\hat{z}}, M_{\hat{z},I_i} = m \} \cap \{S_{2,i} = s_2^*\}\right]
         \nonumber \\
         = & \mathbb{P}\left[\cap_{\hat{z} \in \mathcal{\hat{Z}}} \{S_{1,\hat{z},i} =s_{1,\hat{z}}, M_{\hat{z},I_i} = m \} \right] \mathbb{P} \left[S_{2,i} = s_2^*|\cap_{\hat{z} \in \mathcal{\hat{Z}}} \{S_{1,\hat{z},i} =s_{1,\hat{z}}, M_{\hat{z},I_i} = m \}\right]
         \nonumber \\ 
          = & \mathbb{P}\left[\cap_{\hat{z} \in \mathcal{\hat{Z}}} \{S_{1,\hat{z},i} =s_{1,\hat{z}}, M_{\hat{z},I_i} = m \} \right] \mathbb{P} \left[S_{2,i} = s_2^*|\cap_{\hat{z} \in \mathcal{\hat{Z}}} \{S_{1,\hat{z},i} =s_{1,\hat{z}}\}\right]
         \label{eq:last05} \\ 
         = & C \cdot \mathbb{P}\left[\cap_{\hat{z} \in \mathcal{\hat{Z}}} \{S_{1,\hat{z},i} =s_{1,\hat{z}}, M_{\hat{z},I_i} = m \}\right].
         \label{eq:last1}
\end{align}
where \eqref{eq:last0} follows from~\eqref{eq:dddd}, \eqref{eq:last05} follows from the Markov chain $S_{2,i}-(S_{1,\hat{z},i})_{\hat{z} \in \hat{\mathcal{Z}}}- (M_{\hat{z},I_i})_{\hat{z} \in \hat{\mathcal{Z}}}$, and $C=\mathbb{P} \left[S_{2,i} = s_2^*|\cap_{\hat{z} \in \mathcal{\hat{Z}}} \{S_{1,\hat{z},i} =s_{1,\hat{z}}\}\right]>0$.
Since $(S_1,S_2)$ is $\mathcal{S}_1$-regular, we have $(S_{1,\hat{z}})_{\hat{z} \in \mathcal{\hat{Z}}}$ has a full support. By \Cref{lem:rehy}, we have
\begin{align}
    &\mathbb{P}\left[\cap_{\hat{z} \in \mathcal{\hat{Z}}} \{S_{1,\hat{z},i} =s_{\hat{z},1}, M_{\hat{z},I_i} = m \}\right] 
     \nonumber \\ 
     \geq &
     \prod_{\hat{z} \in \mathcal{\hat{Z}}}(\mathbb{P}\left[S_{1,\hat{z},i} =s_{\hat{z},1}, M_{\hat{z},I_i} = m \right])^{\alpha_{\hat{z}}}
     \nonumber \\
     = &
     \prod_{\hat{z} \in \mathcal{\hat{Z}}}(\mathbb{P}\left[S_{1,i} =s_{\hat{z},1}, M_{I_i} = m \right])^{\alpha_{\hat{z}}},
     \label{eq:last2}
 \end{align}
where $\alpha_{\hat{z}}>1$.
Then by~\eqref{eq:lowerprob}, we have
\begin{align*}
&\mathbb{P}\left[S_{1,i} =s_{\hat{z},1}, M_{I_i} 
= m \right]
\\
\geq &\mathbb{P}[ M_{I_i} = m,S_{1,i} =s_{1,\hat{z}},S_{2,i} = s_{2}^*] 
\\
\geq &\frac{1}{|\mathcal{S}_1||\mathcal{S}_2|}2^{-t}\delta,
\end{align*}
which together with~\eqref{eq:last1} and ~\eqref{eq:last2} implies 
\begin{align*}
         \mathbb{P}[d(Z_i,\hat{Z}_i)> \epsilon] \geq C \prod_{\hat{z} \in \mathcal{\hat{Z}}}\left(\frac{1}{|\mathcal{S}_1||\mathcal{S}_2|}2^{-t}\delta\right)^{\alpha_{\hat{z}}} \triangleq C'>0,
\end{align*}
and $C'$ does not depend on $n$.
This contradicts to the achivability~\eqref{eq:achivable} of the $(n,2^{nR},t)$ code, and thus $R \geq  \tilde{R}_{loc}(\epsilon)$, which completes the proof.

%% file: A5.tex
\section{Proof of Theorem~\ref{thm:sideinformationgraph}}
\label{pfthm:sideinformationgraph}
By~\eqref{eq:R(e)}, we need to show that 
\begin{equation}
\min_{\substack{p(\tilde{u}|s_1):\exists g, \\  \mathbb{P}[d(Z,g(\tilde{U},S_2))\leq \epsilon]=1}} I(\tilde{U};S_1) = \min_{\substack{p(u|s_1):\\(S_1,U) \in \mathcal{E}_{\epsilon}}} I(U;S_1).
\end{equation}

{\it Step 1.} We first prove ``$\leq$".
Suppose that $p(u|s_1)$
satisfies $(S_1,U) \in \mathcal{E}_{\epsilon}$. 
By the definition of $\Gamma_{\epsilon}$, for any  $u \in \Gamma_{\epsilon}$ and $s_2 \in \mathcal{S}_2$, there exists some $\hat{z}$ such that for any $s_1 \in u$ with $p(s_1,s_2)>0$, we have $d(f(s_1,s_2),\hat{z}) \leq \epsilon$. Then we let $g(u,s_2) = \hat{z}$.

For any $(u,s_1,s_2)$ with $p(u,s_1,s_2)>0$, we have $p(u|s_1)>0$ and $p(s_1,s_2)>0$. 
Then we have $(s_1,u) \in \mathcal{E}_{\epsilon}$, so by the definition of $\mathcal{E}_{\epsilon}$, we have $s_1 \in u$.
Therefore, by the definition of $g$ we have $d(f(s_1,s_2),g(u,s_2)) \leq \epsilon$. This shows $\mathbb{P}[d(Z,g(U,S_2))\leq \epsilon]=1$.
Then $p(u|s_1)$ and $g$ defined above satisfy constraints in the left side, which proves  ``$\leq$”.

{\it Step 2.} Next we show the other direction of ``$\geq$". 
Let $p(\tilde{u}|s_1)$ and $g$ satisfy $\mathbb{P}[d(f(S_1,S_2),g(\tilde{U},S_2)) \leq \epsilon] = 1$.
For any $\tilde{u} \in \tilde{\mathcal{U}}$, define a set 
\begin{equation*}
u(\tilde{u}) = \{ s_1 \in \mathcal{S}_1: p(\tilde{u}|s_1) > 0 \}, 
\end{equation*}
and let $U=u(\tilde{U})$, which implies $S_1 \in U$ almost surely.  We see that $U$ is a function of $\tilde{U}$, which implies the Markov chain $U-\tilde{U}-S_1-S_2$.
Since $\mathbb{P}[d(f(S_1,S_2),g(\tilde{U},S_2)) \leq \epsilon] = 1$, for any $u(\tilde{u}) \in u(\tilde{\mathcal{U}})$ and $s_2 \in \mathcal{S}_2$, $g(\tilde{u},s_2)$ satisfies that, for any $s_1 \in u(\tilde{u})$ (which implies that $p(\tilde{u}|s_1)>0$) with $p(s_1,s_2)>0$, we have $d(f(s_1,s_2),g(\tilde{u},s_2)) \leq \epsilon$. By~\Cref{def:graphsideinformation}, we have $u(\tilde{\mathcal{U}}) \subseteq \Gamma_{\epsilon}$, thus we have $U \in \Gamma_{\epsilon}$ and $(S_1,U) \in \mathcal{E}_{\epsilon}$ almost surely.
Hence $p(u|s_1)$ satisfies the constraint on the right side.
Moreover we have
$I(\tilde{U};S_1)\geq I(U;S_1)$ by the data processing inequality. 
This completes the proof. 

%% file: A6.tex
\section{Proof of Theorem~\ref{thm:distributedgraph}}
\label{pfthm:distributedgraph}
By~\eqref{eq:innerbound}, the achievable bound can be expressed as $\tilde{\mathcal{R}}_{loc}(\epsilon) = \mathrm{conv} (E \cup \{(\infty,\infty)\})$, where $E$ is 
the set of rate pairs $(R_1,R_2)$ such that 
\begin{equation}
\begin{aligned}
    R_1 = I(\tilde{U}_1;X_1),
    \\
    R_2 = I(\tilde{U}_2;X_2),
\end{aligned}
\end{equation}
for some conditional distribution $p(\tilde{u}_1|x_1)p(\tilde{u}_2|x_2)$ and function $g:\tilde{\mathcal{U}}_1 \times \tilde{\mathcal{U}}_2 \to \hat{Z}$ such that $\mathbb{P}[d(Z,g(\tilde{U}_1,\tilde{U}_2)) \leq \epsilon] = 1$.
Then we need to prove that 
\begin{equation}
\mathrm{conv} (E \cup \{(\infty,\infty)\}) = \mathrm{conv} (E_{\mathcal{G}} \cup \{(\infty,\infty)\}).
\end{equation}
It suffices to show two directions $E_{\mathcal{G}} \subseteq \mathrm{conv} (E \cup \{(\infty,\infty)\})$ and $E \subseteq \mathrm{conv} (E_{\mathcal{G}} \cup \{(\infty,\infty)\})$.

{\it Step 1.} We first prove ``$E_{\mathcal{G}} \subseteq \mathrm{conv} (E \cup \{(\infty,\infty)\})$" by showing $E_{\mathcal{G}} \subseteq E$.
Suppose that we have a characteristic bipartite graph $\mathcal{G}_{D}$ consisting of $\mathcal{G}_1[\mathcal{X}_1,\Gamma_{1,\epsilon},\mathcal{E}_{1,\epsilon}]$ and $\mathcal{G}_2[\mathcal{X}_2,\Gamma_{2,\epsilon},\mathcal{E}_{2,\epsilon}]$ such that
\begin{align*}
    R_1  = \min_{p(u_1|x_1):(X_1, U_1) \in \mathcal{E}_{1,\epsilon}} I(X_1;U_1),
    \\
    R_2 = \min_{p(u_2|x_2):(X_2, U_2) \in \mathcal{E}_{2,\epsilon}} I(X_2;U_2).
\end{align*}
Since $\mathcal{G}_D$ is a distributed characteristic bipartite graph, for any  $u_1 \in \Gamma_{1,\epsilon}$ and $u_2 \in \Gamma_{2,\epsilon}$, there exists some $\hat{z} \in \hat{\mathcal{Z}}$ such that for any $x_1 \in u_1$ and $x_2 \in u_2$ with $p(x_1,x_2)>0$, we have $d(f(x_1,x_2),\hat{z}) \leq \epsilon$. Then we let $g(u_1,u_2) = \hat{z}$.

For any $(u_1,,u_2,x_1,x_2)$ with $p(u_1,u_2,x_1,x_2)>0$, we have $p(u_1|x_1)>0$, $p(u_2|x_2)>0$ and $p(x_1,x_2)>0$. 
Then we have $(x_k,u_k) \in \mathcal{E}_{k,\epsilon}$, so by the definition of $\mathcal{E}_{k,\epsilon}$, we have $s_k \in u_k$, $k = 1,2$.
Therefore, by the definition of $g$ we have $d(f(x_1,x_2),g(u_1,u_2)) \leq \epsilon$. This shows $\mathbb{P}[d(Z,g(U_1,U_2))\leq \epsilon]=1$.
Then $p(u_1|x_1)p(u_2|x_2)$ and $g$ defined above satisfy constraints for rate pairs $(R_1,R_2)$ to be in $E$, which proves  $E_{\mathcal{G}} \subseteq E$.

{\it Step 2.} We show ``$E \subseteq \mathrm{conv} (E_{\mathcal{G}} \cup \{(\infty,\infty)\})$". 
Let $(\tilde{R}_1.\tilde{R}_2) \in E$, then $\tilde{R}_k = I(\tilde{U}_k;X_k)$, $k = 1,2$, for some conditional distribution $p(\tilde{u}_1|x_1)p(\tilde{u}_2|x_2)$ and function $g$ such that $\mathbb{P}[d(f(X_1,X_2),g(\tilde{U}_1,\tilde{U}_2)) \leq \epsilon] = 1$.
For any $k = 1,2$ and $\tilde{u}_k \in \tilde{\mathcal{U}}_k$, define a set 
\begin{equation*}
u_k(\tilde{u}_k) = \{ x_k \in \mathcal{X}_k: p(\tilde{u}_k|x_k) > 0 \}, 
\end{equation*}
and let $U_k=u_k(\tilde{U}_k)$, which implies $X_k \in U_k$ almost surely. 
We see that $U_k$ is a function of $\tilde{U}_k$, which implies the Markov chain $U_1-\tilde{U}_1-X_1-X_2-\tilde{U}_2-U_2$.
Then let $\Gamma_{k,\epsilon} = u_k(\tilde{\mathcal{U}}_k)$ and $\mathcal{E}_{k,\epsilon} = \{(x_k,u_k)|x_k \in u_k \in \Gamma_{k,\epsilon}\}$, $k = 1,2$. We can construct $\mathcal{G}_{D}$ from two bipartite graphs $\mathcal{G}_1[\mathcal{X}_1,\Gamma_{1,\epsilon},\mathcal{E}_{1,\epsilon}]$ and $\mathcal{G}_2[\mathcal{X}_2,\Gamma_{2,\epsilon},\mathcal{E}_{2,\epsilon}]$.

Next we show $\mathcal{G}_{D}$ is a distributed characteristic bipartite graph. For any $k = 1,2$ and $x_k \in \mathcal{X}_k$, there exists some $\tilde{u}_k$ with $p(\tilde{u}_k|x_k)>0$, hence $x_k \in u_k(\tilde{u}_k)$, where $u_k(\tilde{u}_k) \in \Gamma_{k,\epsilon}$.
Since $\mathbb{P}[d(f(X_1,X_2),g(\tilde{U}_1,\tilde{U}_2)) \leq \epsilon] = 1$, for any $u_k(\tilde{u}_k) \in u_k(\tilde{\mathcal{U}}_k)$, $k = 1,2$, $g(\tilde{u}_1,\tilde{u}_2)$ satisfies that, for any $x_k \in u_k(\tilde{u}_k)$, $k = 1,2$, (which implies that $p(\tilde{u}_k|x_k)>0$) with $p(x_1,x_2)>0$, we have $d(f(x_1,x_2),g(\tilde{u}_k,\tilde{u}_2)) \leq \epsilon$. Therefore by~\Cref{def:graphdistributed}, $\mathcal{G}_{D}$ is a distributed characteristic bipartite graph.

By the definition of $\Gamma_{k,\epsilon}$ and $\mathcal{E}_{k,\epsilon}$, we have $U_k \in \Gamma_{k,\epsilon}$ and $(X_k,U_k) \in \mathcal{E}_{k,\epsilon}$ almost surely for $k = 1,2$.
Hence $p(u_k|x_k)$ satisfies the constraint of the minimization problem in~\eqref{eq:graphinnerbound}.
Moreover we have
$\tilde{R}_k = I(\tilde{U}_k;X_k)\geq I(U_k;X_k)$ by the data processing inequality. 
Since $(I(U_1;X_1),I(U_2;X_2)) \in E_{\mathcal{G}}$, we have  $(\tilde{R}_1,\tilde{R}_2) \in \mathrm{conv} (E_{\mathcal{G}} \cup \{(\infty,\infty)\})$. Hence $E \subseteq \mathrm{conv} (E_{\mathcal{G}} \cup \{(\infty,\infty)\})$, which completes the proof. 

%% file: main.bbl
\begin{thebibliography}{10}
\providecommand{\url}[1]{#1}
\csname url@samestyle\endcsname
\providecommand{\newblock}{\relax}
\providecommand{\bibinfo}[2]{#2}
\providecommand{\BIBentrySTDinterwordspacing}{\spaceskip=0pt\relax}
\providecommand{\BIBentryALTinterwordstretchfactor}{4}
\providecommand{\BIBentryALTinterwordspacing}{\spaceskip=\fontdimen2\font plus
\BIBentryALTinterwordstretchfactor\fontdimen3\font minus
  \fontdimen4\font\relax}
\providecommand{\BIBforeignlanguage}[2]{{%
\expandafter\ifx\csname l@#1\endcsname\relax
\typeout{** WARNING: IEEEtran.bst: No hyphenation pattern has been}%
\typeout{** loaded for the language `#1'. Using the pattern for}%
\typeout{** the default language instead.}%
\else
\language=\csname l@#1\endcsname
\fi
#2}}
\providecommand{\BIBdecl}{\relax}
\BIBdecl

\bibitem{Soyata2012}
T.~Soyata, R.~Muraleedharan, C.~Funai, M.~Kwon, and W.~Heinzelman,
  ``Cloud-{Vision}: Real-time face recognition using a mobile-cloudlet-cloud
  acceleration architecture,'' in \emph{2012 IEEE Symposium on Computers and
  Communications}, Cappadocia, Turkey, Jul. 2012, pp. 59--66.

\bibitem{Nedic2009}
A.~Nedic and A.~Ozdaglar, ``Distributed subgradient methods for multi-agent
  optimization,'' \emph{IEEE Transactions on Automatic Control}, vol.~54,
  no.~1, pp. 48--61, Jan. 2009.

\bibitem{Lee2018}
K.~Lee, M.~Lam, R.~Pedarsani, D.~Papailiopoulos, and K.~Ramchandran, ``Speeding
  up distributed machine learning using codes,'' \emph{IEEE Transactions on
  Information Theory}, vol.~64, no.~3, pp. 1514--1529, Mar. 2018.

\bibitem{Yamamoto1982}
H.~Yamamoto, ``Wyner-{Ziv} theory for a general function of the correlated
  sources,'' \emph{IEEE Transactions on Information Theory}, vol.~28, no.~5,
  pp. 803--807, Sep. 1982.

\bibitem{Orlitsky2001}
A.~Orlitsky and J.~Roche, ``Coding for computing,'' \emph{IEEE Transactions on
  Information Theory}, vol.~47, no.~3, pp. 903--917, Mar. 2001.

\bibitem{Han1987}
T.~Han and K.~Kobayashi, ``A dichotomy of functions {$F(X,Y)$} of correlated
  sources {$(X, Y)$},'' \emph{IEEE Transactions on Information Theory},
  vol.~33, no.~1, pp. 69--76, Jan. 1987.

\bibitem{SlepianWolf1973}
D.~Slepian and J.~Wolf, ``Noiseless coding of correlated information sources,''
  \emph{IEEE Transactions on Information Theory}, vol.~19, no.~4, pp. 471--480,
  Jul. 1973.

\bibitem{WynerZiv1976}
A.~Wyner and J.~Ziv, ``The rate-distortion function for source coding with side
  information at the decoder,'' \emph{IEEE Transactions on Information Theory},
  vol.~22, no.~1, pp. 1--10, Jan. 1976.

\bibitem{Berger1978}
T.~Berger, ``Multiterminal source coding,'' in \emph{The Information Theory
  Approach to Communications}, G.~Longo, Ed.\hskip 1em plus 0.5em minus
  0.4em\relax New York, NY, USA: Springer-Verlag, 1978, pp. 171--231.

\bibitem{Tung1978}
S.-Y. Tung, ``Multiterminal source coding,'' Ph.D. dissertation, School of
  Electrical Engineering, Cornell University, Ithaca, NY, USA, May 1978.

\bibitem{Wagner20081}
A.~B. Wagner and V.~Anantharam, ``An improved outer bound for multiterminal
  source coding,'' \emph{IEEE Transactions on Information Theory}, vol.~54,
  no.~5, pp. 1919--1937, May 2008.

\bibitem{KornerMarton1979}
J.~Korner and K.~Marton, ``How to encode the modulo-two sum of binary
  sources,'' \emph{IEEE Transactions on Information Theory}, vol.~25, no.~2,
  pp. 219--221, Mar. 1979.

\bibitem{Wagner20082}
A.~B. Wagner, S.~Tavildar, and P.~Viswanath, ``Rate region of the quadratic
  gaussian two-encoder source-coding problem,'' \emph{IEEE Transactions on
  Information Theory}, vol.~54, no.~5, pp. 1938--1961, May 2008.

\bibitem{Xu2016}
Q.~Xu, T.~Mytkowicz, and N.~S. Kim, ``Approximate computing: A survey,''
  \emph{IEEE Design {\&} Test}, vol.~33, no.~1, pp. 8--22, Feb. 2016.

\bibitem{Posner1971}
E.~C. Posner and E.~R. Rodemich, ``{Epsilon Entropy and Data Compression},''
  \emph{The Annals of Mathematical Statistics}, vol.~42, no.~6, pp. 2079 --
  2125, Dec. 1971.

\bibitem{Basu2020}
S.~Basu, D.~Seo, and L.~R. Varshney, ``Functional epsilon entropy,'' in
  \emph{2020 Data Compression Conference}, Snowbird, UT, USA, Mar. 2020, pp.
  332--341.

\bibitem{Basu2022}
------, ``Hypergraph-based source codes for function computation under maximal
  distortion,'' \emph{IEEE Journal on Selected Areas in Information Theory},
  vol.~3, no.~4, pp. 824--838, Dec. 2022.

\bibitem{Marton1974}
K.~Marton, ``Error exponent for source coding with a fidelity criterion,''
  \emph{IEEE Transactions on Information Theory}, vol.~20, no.~2, pp. 197--199,
  Mar. 1974.

\bibitem{Doshi2010}
V.~Doshi, D.~Shah, M.~Médard, and M.~Effros, ``Functional compression through
  graph coloring,'' \emph{IEEE Transactions on Information Theory}, vol.~56,
  no.~8, pp. 3901--3917, Aug. 2010.

\bibitem{Yuan2022}
D.~Yuan, T.~Guo, B.~Bai, and W.~Han, ``Lossy computing with side information
  via multi-hypergraphs,'' in \emph{2022 IEEE Information Theory Workshop},
  Mumbai, India, Nov. 2022, pp. 344--349.

\bibitem{Yuan2023}
\BIBentryALTinterwordspacing
D.~Yuan, T.~Guo, and Z.~Huang, ``A graph-based approach to the computation of
  rate-distortion and capacity-cost functions with side information,'' 2023.
  [Online]. Available: \url{https://arxiv.org/abs/2306.04981}
\BIBentrySTDinterwordspacing

\bibitem{Makhdoumi2015}
A.~Makhdoumi, S.-L. Huang, M.~Médard, and Y.~Polyanskiy, ``On locally
  decodable source coding,'' in \emph{2015 IEEE International Conference on
  Communications}, London, UK, Jun. 2015, pp. 4394--4399.

\bibitem{Mazumdar2014}
A.~Mazumdar, V.~Chandar, and G.~W. Wornell, ``Update-efficiency and local
  repairability limits for capacity approaching codes,'' \emph{IEEE Journal on
  Selected Areas in Communications}, vol.~32, no.~5, pp. 976--988, May 2014.

\bibitem{Montanari2008}
A.~Montanari and E.~Mossel, ``Smooth compression, {Gallager} bound and
  nonlinear sparse-graph codes,'' in \emph{2008 IEEE International Symposium on
  Information Theory}, Toronto, ON, Canada, Jul. 2008, pp. 2474--2478.

\bibitem{Vatedka2020}
S.~Vatedka and A.~Tchamkerten, ``Local decode and update for big data
  compression,'' \emph{IEEE Transactions on Information Theory}, vol.~66,
  no.~9, pp. 5790--5805, Sep. 2020.

\bibitem{Mazumdar2021}
A.~Mazumdar and S.~Pal, ``Semisupervised clustering by queries and locally
  encodable source coding,'' \emph{IEEE Transactions on Information Theory},
  vol.~67, no.~2, pp. 1141--1155, Feb. 2021.

\bibitem{Mazumdar2015}
A.~Mazumdar, V.~Chandar, and G.~W. Wornell, ``Local recovery in data
  compression for general sources,'' in \emph{2015 IEEE International Symposium
  on Information Theory}, Hong Kong, China, Jun. 2015, pp. 2984--2988.

\bibitem{Pananjady2018}
A.~Pananjady and T.~A. Courtade, ``The effect of local decodability constraints
  on variable-length compression,'' \emph{IEEE Transactions on Information
  Theory}, vol.~64, no.~4, pp. 2593--2608, Apr. 2018.

\bibitem{Tatwawadi2018}
K.~Tatwawadi, S.~S. Bidokhti, and T.~Weissman, ``On universal compression with
  constant random access,'' in \emph{2018 IEEE International Symposium on
  Information Theory}, Vail, CO, USA, Jun. 2018, pp. 891--895.

\bibitem{Buhrman2002}
H.~Buhrman, P.~B. Miltersen, J.~Radhakrishnan, and S.~Venkatesh, ``Are
  bitvectors optimal?'' \emph{SIAM Journal on Computing}, vol.~31, no.~6, pp.
  1723--1744, Jun. 2002.

\bibitem{Vatedka2022}
S.~Vatedka, V.~Chandar, and A.~Tchamkerten, ``Local decoding in distributed
  compression,'' \emph{IEEE Journal on Selected Areas in Information Theory},
  vol.~3, no.~4, pp. 711--719, Dec. 2022.

\bibitem{EIGamal2011}
A.~El~Gamal and Y.-H. Kim, \emph{Network Information Theory}.\hskip 1em plus
  0.5em minus 0.4em\relax Cambridge: Cambridge University Press, 2011.

\bibitem{Mossel2013}
E.~Mossel, K.~Oleszkiewicz, and A.~Sen, ``On reverse hypercontractivity,''
  \emph{Geometric and Functional Analysis}, vol.~23, no.~3, p. 1062–1097,
  Jun. 2013.

\bibitem{Weissman2006}
T.~Weissman and A.~El~Gamal, ``Source coding with limited-look-ahead side
  information at the decoder,'' \emph{IEEE Transactions on Information Theory},
  vol.~52, no.~12, pp. 5218--5239, Dec. 2006.

\bibitem{Korner1973}
J.~K\"{o}rner, ``Coding of an information source having ambiguous alphabet and
  the entropy of graphs,'' in \emph{6th Prague Conference on Information
  Theory, etc.}, Prague, Czech, Sep. 1973, pp. 411--425.

\end{thebibliography}
